\documentclass[aip,graphicx]{revtex4-1}

\usepackage{latexsym,pslatex,psfrag,times,color,amsmath}
\usepackage[latin1]{inputenc}
\usepackage{graphicx}
\usepackage{subfigure}
\usepackage{setspace}

\begin{document}

\title{Structure-dependent mobility of a dry aqueous foam flowing along two parallel channels}

\author{S. A. Jones}
\affiliation{Institut de Physique de Rennes (UMR CNRS 6251), Universit\'e de Rennes 1, 35042  Rennes, France}
\author{B. Dollet}
\affiliation{Institut de Physique de Rennes (UMR CNRS 6251), Universit\'e de Rennes 1, 35042  Rennes, France}
\author{Y. M\'eheust}
\affiliation{G\'eosciences (UMR CNRS 6118),  Universit\'e de Rennes 1, 35042 Rennes, France}
\author{S. J.  Cox}
\affiliation{Institute of Mathematics and Physics, Aberystwyth University, Aberystwyth SY23 1BN, UK}
\author{I. Cantat}
\affiliation{Institut de Physique de Rennes (UMR CNRS 6251), Universit\'e de Rennes 1, 35042 Rennes, France}

\date{Version 5}

\begin{abstract}
The velocity of a two-dimensional aqueous foam has been measured as it flows
through two parallel channels, at a constant overall volumetric flow rate.
The flux distribution between the two channels is studied as a function of
the ratio of their widths.  A peculiar dependence of the velocity
ratio on the width ratio is observed when the foam structure in the narrower channel is
either single staircase or bamboo. In particular, discontinuities in the
velocity ratios are observed at the transitions between double and single staircase
and between single staircase and bamboo. A theoretical model accounting for the viscous
dissipation at the solid wall and the capillary pressure across a film pinned at the
channel outlet  predicts the observed non-monotonic evolution of the velocity ratio as a
function of the width ratio. It also predicts quantitatively the intermittent
temporal evolution of the velocity in the narrower channel when it is so narrow that film pinning at
its outlet repeatedly brings the flow to a near stop.
\end{abstract}

\maketitle

\section{Introduction}

Aqueous foams are present in many manufactured goods, from personal hygiene
products to processed foods, but also play an important role in many industrial
processes \cite{weaire,livre_mousse}. Some of these processes involve the
injection of a foam into the Earth's surface or subsurface, in particular to
enhance oil recovery \cite{rossen96} or as a carrier fluid for soil remediation
processes \cite{wang04}. This latter application is recent and particularly
promising as there are many expected advantages of injecting a foam rather than a
single phase fluid in the polluted soil.  
It allows a large reduction of the
needed volume of liquid for a given injection volume, while maintaining a very
good compatibility with the surfactants already used for soil remediation
\cite{mulligan01,mulligan03}.  There is also the  potential capacity for the selective
transport of material by the foam, in the form of particles of polluted soil or
colloidal pollutant, in the same manner as in ore separation by froth flotation.
In addition, if
bioremediation of a soil is required, the transport of air along with the liquid
in a foam may enhance the efficiency of any biological activity
\cite{rothmel98}.

In both enhanced oil recovery and soil remediation a foam flows  
through a porous material with a complex geometry. Being
able to predict how a foam flows in a confined, tortuous, geometry is therefore of great
importance, and has been studied for many years
\cite{hirasaki85,rossen90a,kovscek93,kornev99}.
Laboratory-scale experiments most commonly study the flow through a cylindrical column of a porous material,
consisting of either a model system, such as packed glass beads, or a sample of soil, sand or
rock from the field.

Bertin {\it et al.} \cite{bertin99} obtained a puzzling experimental result
during testing of a heterogeneous porous medium: the (interstitial) velocity of the foam front
invading a saturated porous medium was larger in the low permeability region
than in the high permeability region. The mobility of the foam is classically
quantified through an {\em effective viscosity}, defined as the viscosity of the
Newtonian fluid that would flow at the same velocity in the same geometry under an identical
pressure head \cite{hirasaki85}. For a bubble size larger than the
characteristic pore size, even in heterogeneous porous media, the foam is
organized in trains of lamellae \cite{hirasaki85,rossen90a}. In this case,
Kovscek and Bertin \cite{kovscek03} have shown theoretically that the specific dissipation
associated with this structure leads to an effective viscosity
of the foam that scales with the permeability $K$ of the porous medium  as a power law with an exponent
$3/2$. This key ingredient, also observed in \cite{falls88}, is  at the origin
of the faster invasion of the  low permeability regions of heterogeneous
porous media when foam is used instead of a Newtonian fluid. However, the
complex models developed to address this type of behaviour \cite{kovscek03,du11}
involve many different assumptions about local processes (local dissipation
laws, local foam structure, bubble production and coalescence rate). Full visualisations of
foam flows in model porous media are therefore urgently needed to validate
and improve these theoretical approaches.

Two-dimensional diphasic flows in porous media have been investigated extensively in the last 25 years, allowing researchers to map out the rich phenomenology of flow regimes \cite{lenormand85,lenormand88,maloy85,meheust02}, to finely characterize their dynamics \cite{maloy92,toussaint05}, and gain partial understanding of how the pore scale dynamics results in macroscopic transport properties \cite{toussaint12}. These experimental studies have relied on disordered porous media consisting of networks of linear channels, monolayers of glass beads, or, more recently, microfluidic devices \cite{cottin10}.
 However, the limit in which the 
liquid fraction of the foam is very low has seen little investigation, although bubble formation in this
limit has been investigated in \cite{kovscek07}, for air in pure water or in a
solution of surfactants. This dry foam limit involves specific foam stability issues, making the problem more complicated.

Here we investigate the basic flow behaviour of a dry two-dimensional
foam as it moves through two parallel linear channels enclosed in a larger cell
(see Fig.~\ref{fig:channelheleshaw}).  This simple system mimics the bifurcation
of a large pore into two smaller ones (at the entrance to the two channels), and
the reconnection of two small pores into a larger one (at the exit). The width of each channel is uniform
along its length, and the two widths can be varied systematically for given bubble size and combined channel width. With this
very simple geometry we can investigate how the total foam flux, which is
imposed upstream, is distributed into the two channels. We believe this
represents an important local step toward the understanding of the flux
distributions in heterogeneous porous media.

The pressure drop along such a channel has been measured for a single channel in a
similar geometry \cite{cantat04} and explored theoretically with
the viscous froth  model \cite{kern04,green06,grassia08,raufaste09}. In this study we use these
established results to determine the pressure drop from the velocity and structure of 
the foam, and show that the dissipation law used  in
\cite{kovscek03} is valid for simple trains of lamellae. However, when the foam
structure is more complex, with several bubbles across the
channel section, we show how to modify the model to correctly
reproduce our experimental results. This correction comes from the dependence of the
friction force on the film orientation \cite{cantat04}. We also show that, even for
channels much longer than the bubble size, end effects are important, and they
are much greater for channels venting to open air. We rationalise these end
effects by taking into account the competition between viscous and capillary
forces.

The plan of the paper is as follows.  Our methods are given in \textsection
\ref{methods} and the experimental results for enclosed and open-ended channels
are described in \textsection \ref{results}. In  \textsection \ref{model} we
discuss the theoretical model that we compare to experiments in \textsection
\ref{comp}. Finally, we draw conclusions in \textsection \ref{conclusion}.

\section{Methodology}
\label{methods}
\subsection{Experimental Set-Up}
\label{setup}

\begin{figure}[htb]
\centering
 \subfigure[]{
\includegraphics[width=7cm]{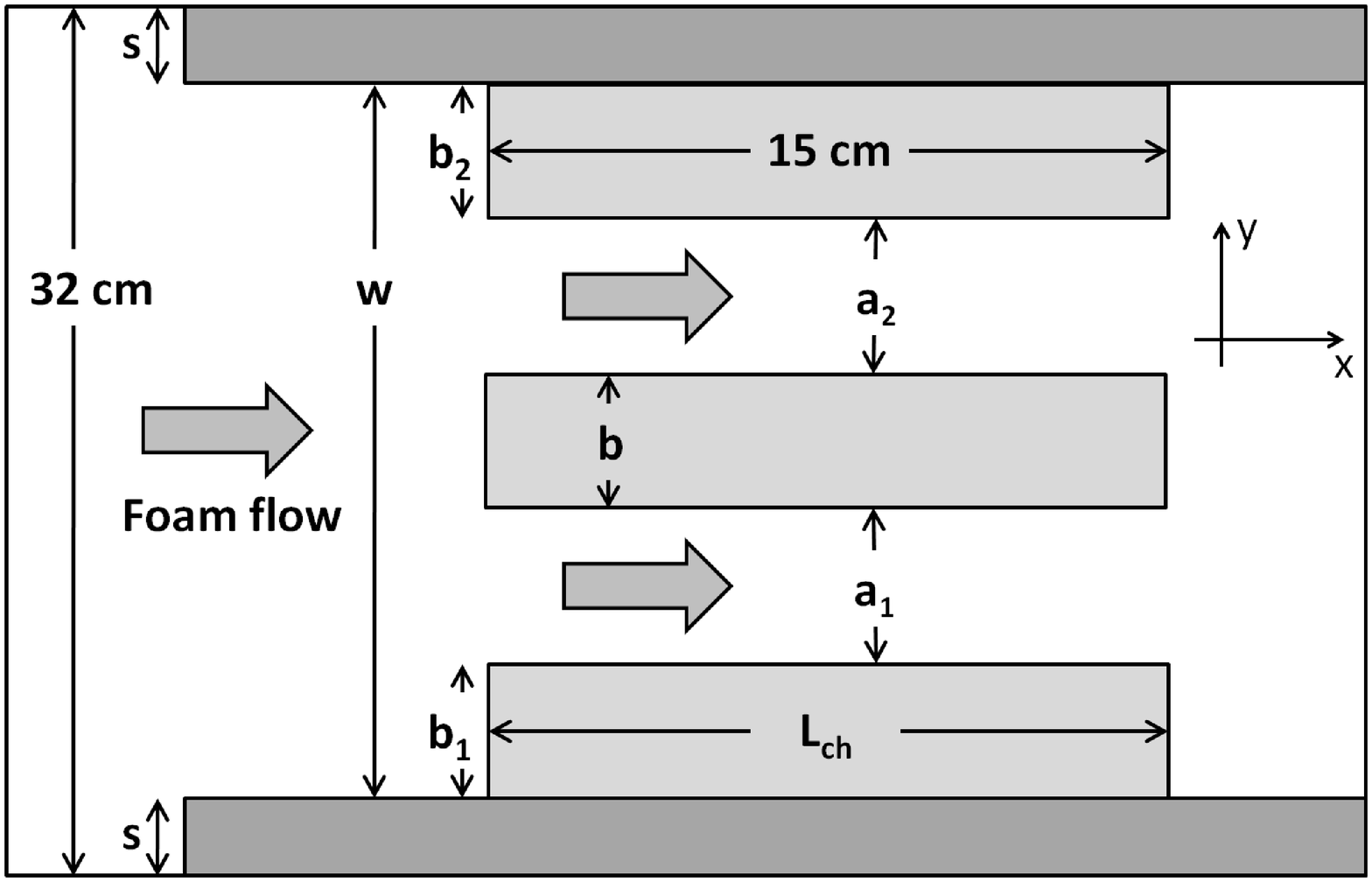}
}
\hspace{0.4cm} 
\subfigure[]{
\includegraphics[width=7cm]{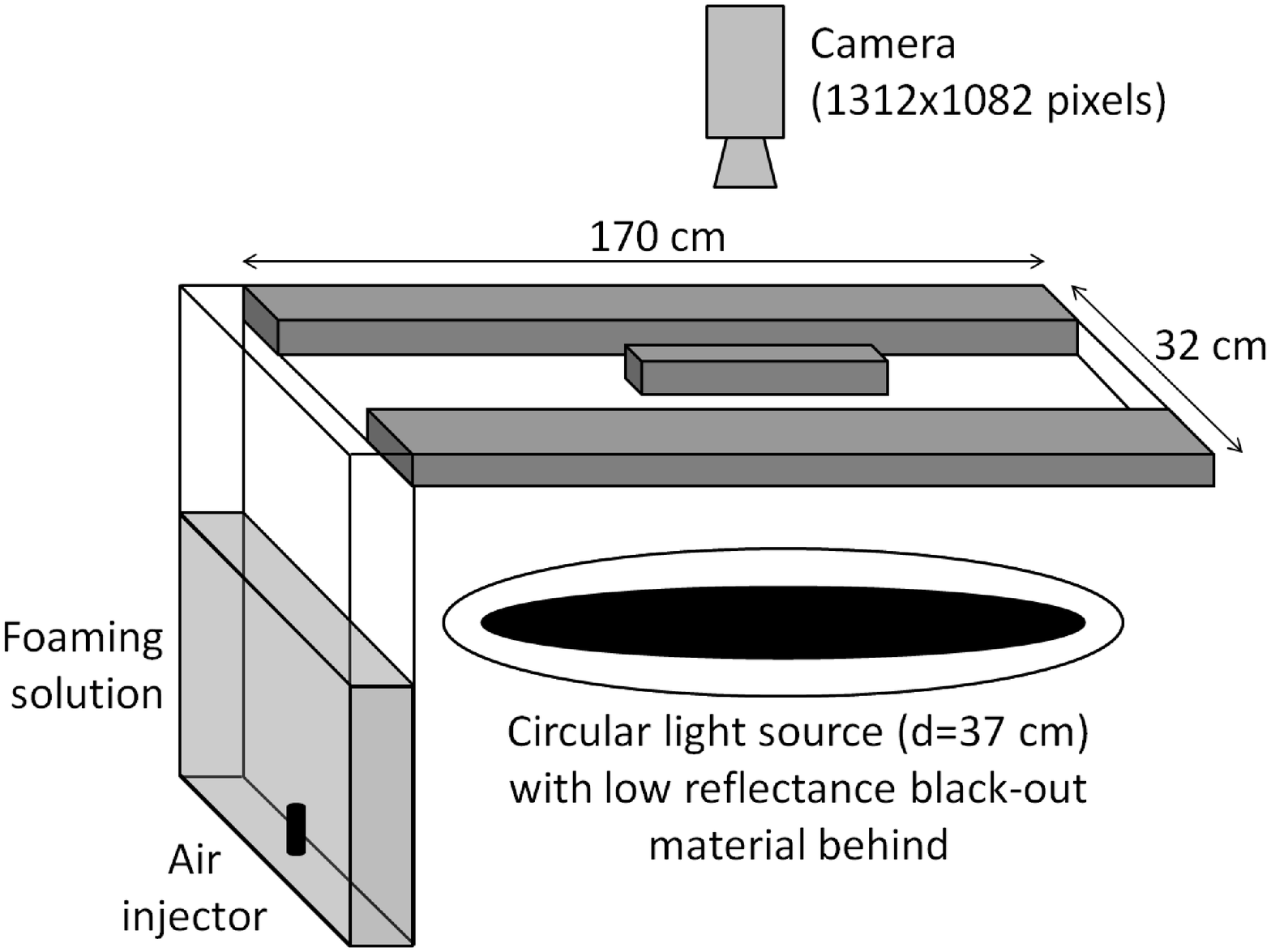}
}
\caption{Schematic representations of (a) the experimental channels, with $a_1$
and $a_2$ the channel widths and $b$ the separation distance between
channels, and (b) the complete experimental setup.}
\label{fig:channelheleshaw}
\end{figure}

The double-channel flow experiments were carried out in a Hele-Shaw cell
with a plate separation of either $h = 0.1$~cm or $h = 0.2$~cm.
Rectangular obstacles cut from polycarbonate sheet were placed within the
Hele-Shaw cell to form two parallel channels.
The channel length $L_\text{ch}$ is 15 cm for all experiments, and the
values of $a_1$, $a_2$, $b$, $b_1$, $b_2$ and $s$, defined on Fig.~\ref{fig:channelheleshaw}(a), could be changed to meet individual test
requirements.
The central obstacle could be moved sideways to study asymmetric channel configurations. 
For the experiment shown in this paper, we  chose  $b_1=b_2=b$ and $(a_1+a_2)/w = 2/5$, unless different values are specified.  The obstacles were
positioned either half-way along the channel, as shown in 
Fig.~\ref{fig:channelheleshaw}(b), so that they were completely surrounded by the
flowing foam (`enclosed channel' configuration) or they could be positioned at
the end of the channel with the foam venting to the atmosphere (`open-ended
channel' configuration).

The foaming solution used in all tests was $10~\rm \text{g/l}$ sodium dodecyl
sulfate (SDS) in ultra-pure water, with surface tension $\gamma = 36.8 \pm
0.3$~mN/m, as measured in \cite{dollet10b}, and viscosity $\mu = 1\times10^{-3}$
Pa$\cdot$s.  Foam was created by blowing nitrogen, at a rate of  $Q=100$~cm$^3$/min, 
through a single nozzle in the base of the
vertical foam production cell (Fig.~\ref{fig:channelheleshaw}(b)).  This produced
a monodisperse foam with a typical bubble volume $\Omega$ deduced from the
contact area $A$ between the bubble and the top glass plate, through the
relation $\Omega=h\, A$.     We define the {\it{equivalent bubble radius}} as
$R= \sqrt{A/\pi}$.

The liquid fraction was estimated from the rate of
decrease of the level of the foaming solution \cite{dollet10b}, 
since the gas flow rate is known. Values in the range $0.01$ to
$0.02$ were obtained for all the tests presented here.

\begin{figure}[htb]
\centerline{
\includegraphics[width=10cm]{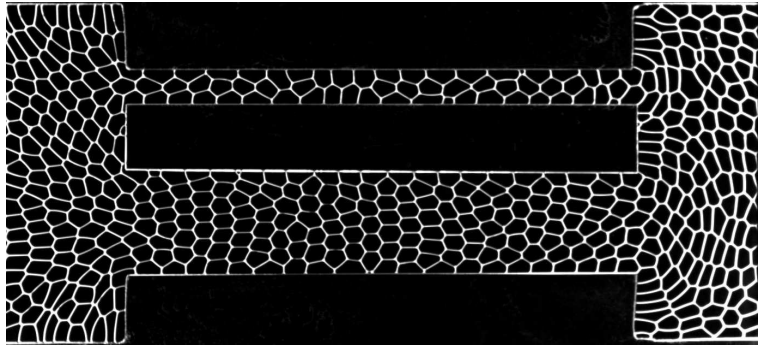}
}
\caption{Image of the foam flow from left to right through two enclosed channels of 
length $15$~cm in a test section of length $22$~cm.}
\label{fig:channelfoam}
\end{figure}

The Hele-Shaw cell was backlit by a circular (diameter~$=37$~cm) fluorescent tube
with a central dark background so that soap films appear white with good contrast.
The motion of the foam was recorded using a $1312\times1082$ pixel digital
video camera running at $25$~fps. A typical image of the foam in the enclosed
channel configuration is given in Fig.~\ref{fig:channelfoam}.

\subsection{Image processing}

Basic image processing, thresholding and skeletonisation of the foam images was
carried out using the ImageJ software package.  The resultant images were then
analysed using an in-house analysis program, described fully in \cite{dollet07}.
 The image treatment procedure tracks individual bubbles to compute the velocity distribution
within the test section over a rectangular mesh of $40\times50$ boxes.  The foam flow in the channels is a plug flow, with a uniform velocity, except in a small domain at the entrance and at the exit. We define the velocities $V_1$ and $V_2$ as the velocities in the central part of the  channels 1 and 2 (outside the entrance and exit domains), averaged over time. The time and space average velocity in the channel of width $w$, before the separation  into two channels, is called $V_0$.

\section{Experimental results}
\label{results}

Two asymmetric channel configurations were investigated: the `enclosed channel'
and the `open-ended channel'. The former  is
relevant to the continuous flow of a foam through a channel within a porous
medium, while the latter is appropriate when considering the penetration of a foam
into an air-filled porous medium, as well as to the flow of a foam along a
channel that emerges into a larger void within a porous medium.

\subsection{Enclosed Channel}

The enclosed channel tests were carried out with a Hele-Shaw plate separation of
$h=0.1$~cm.  
The average bubble area, $A$, was in the range $0.19$ to $0.20$ cm$^2$,
with an equivalent diameter of $2R\approx0.5$~cm and a maximum polydispersity
index of $25\%$.

The first set of tests where performed with channel dimensions of 
$b = b_1 = b_2 = 2$~cm and $a_1 + a_2 = 4$~cm, resulting
in $w = 10$~cm.  The central obstacle was moved from a central position $a_1 =
a_2 = 2$~cm to one side in increments of 
$0.1$ or $0.2$~cm, to give a range of
different $a_1$ and $a_2$, with $a_2$ denoting the width of the narrow
channel. 

\begin{figure}[htb]
\centering
 \subfigure[]{
\includegraphics[width=3.5cm]{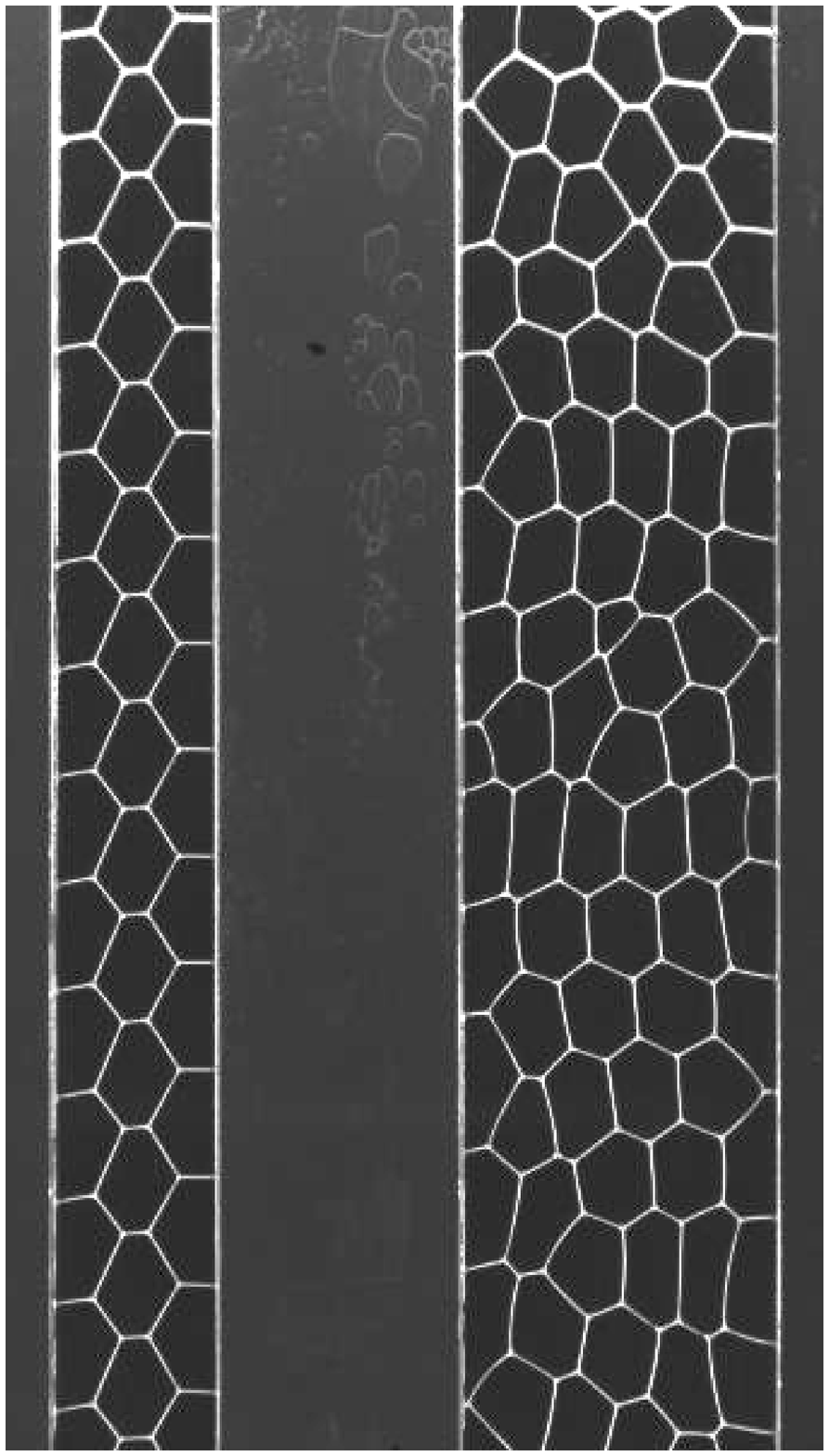}
}
\subfigure[]{
\includegraphics[width=3.5cm]{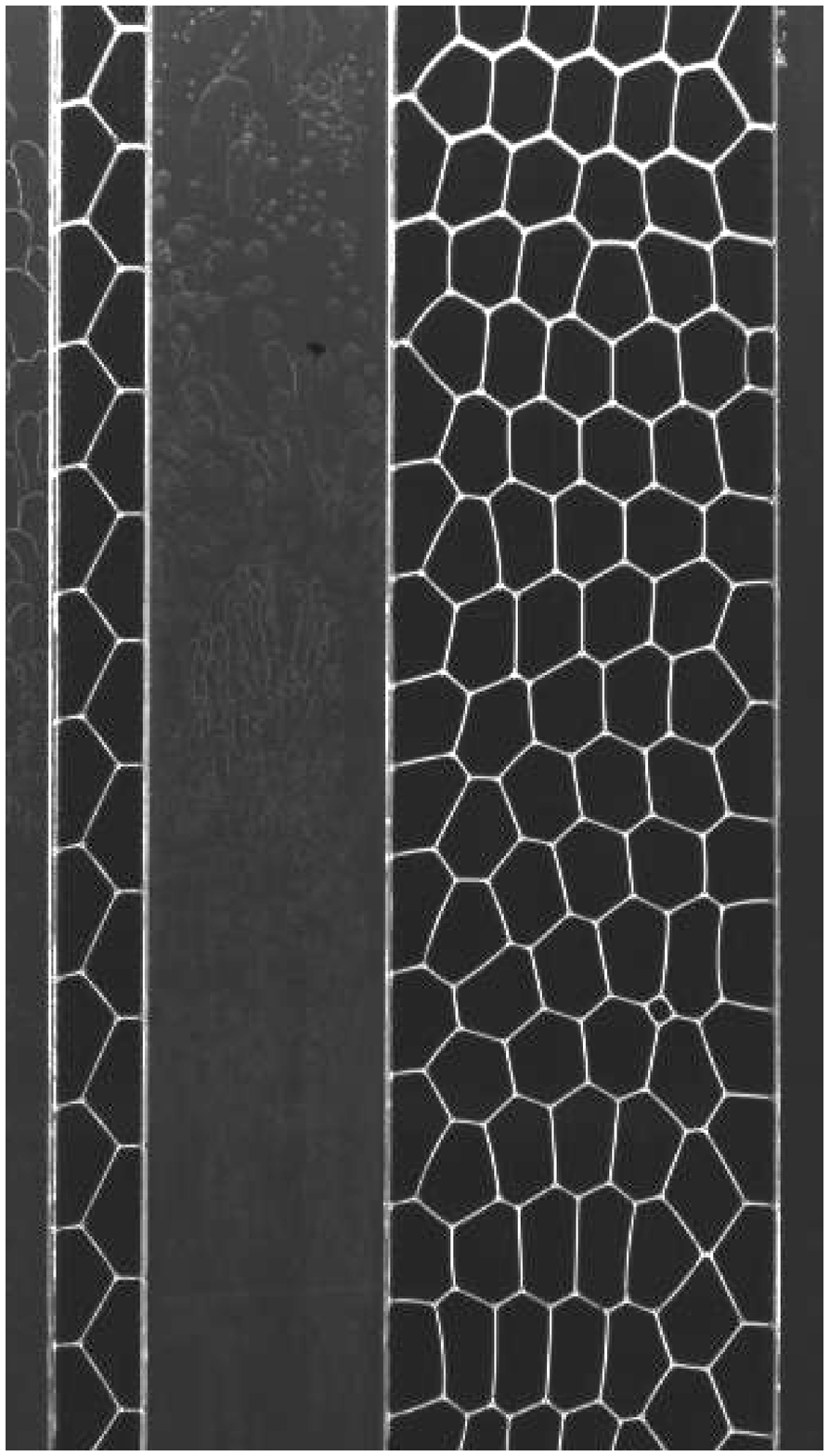}
}
\subfigure[]{
\includegraphics[width=3.5cm]{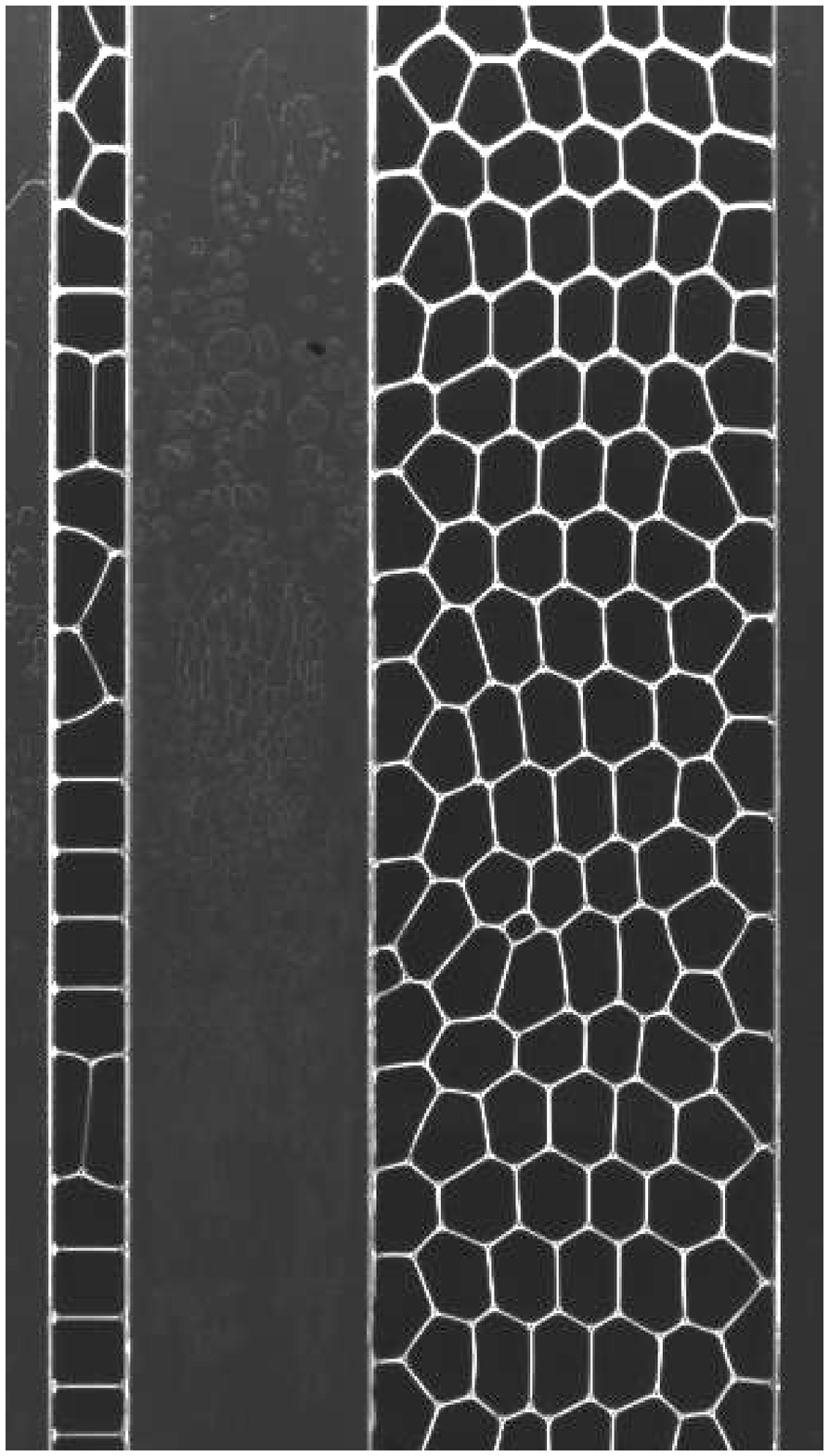}
}
\subfigure[]{
\includegraphics[width=3.5cm]{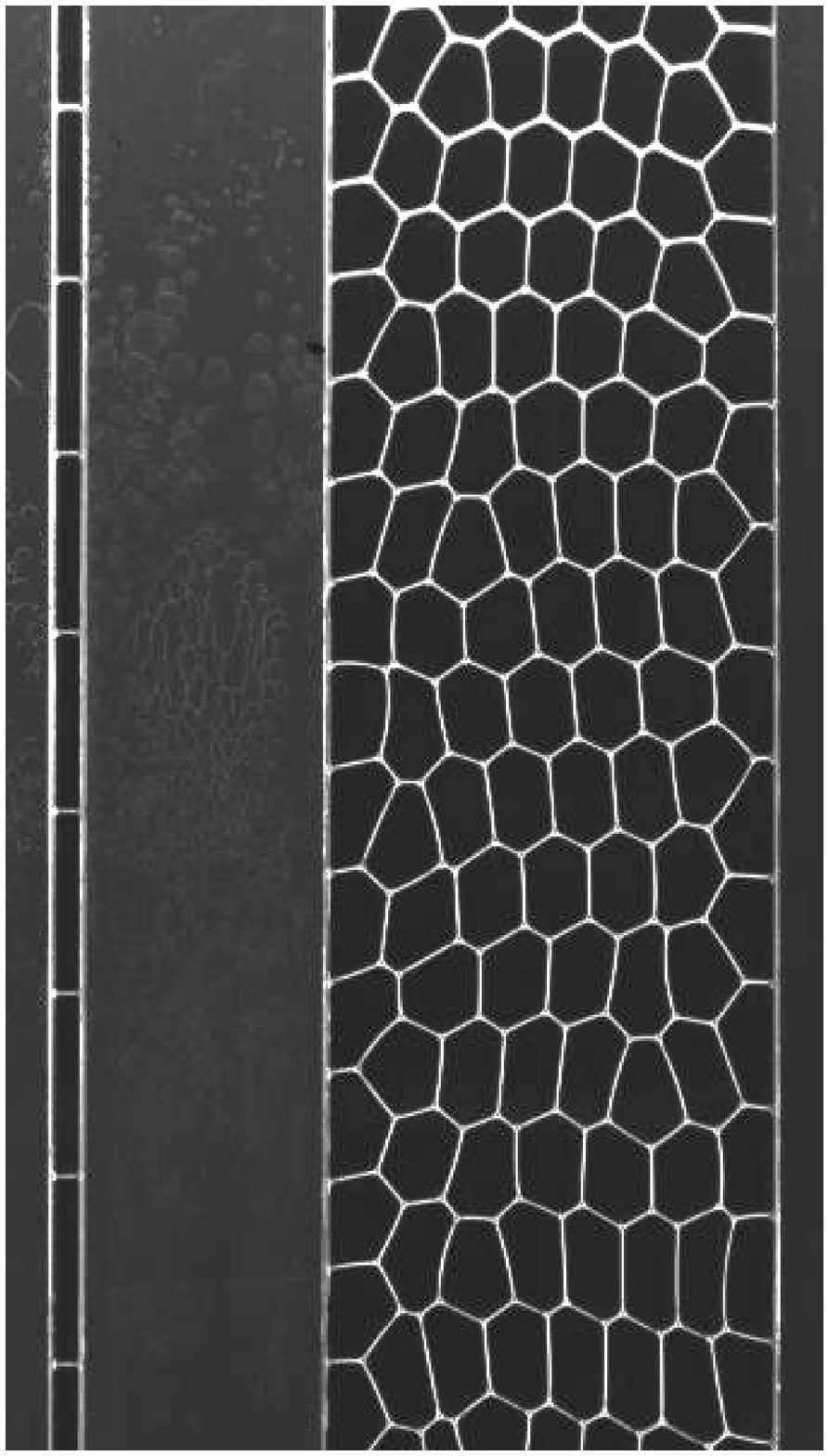}
}
\caption{Images of foam flow through asymmetric enclosed channels
(flow from bottom to top), with widths satisfying $a_1 + a_2 = 4$~cm.  
As the smaller channel becomes narrower, the foam structure
explores a range of ordered and disordered arrangements. Ordered
arrangements are shown for narrow channel widths of (a) $a_2 = 1.33$~cm, (b)
$a_2 = 0.72$~cm,  (c) $a_2 = 0.59$~cm and (d) $a_2 = 0.22$~cm,  resulting in double staircase,
single staircase, mixed staircase/wide bamboo, and narrow bamboo structures,
respectively.  The wider channel contains a random
two-dimensional foam in each case.
}
\label{fig:assymetricfoam}
\end{figure}

Examples of the foam structures observed, with the names that we use for them, 
are shown in Fig.~\ref{fig:assymetricfoam}.
For channel widths much larger than the bubble radius $R$, 
a random foam is observed. For a channel just slightly wider than $2R$, the foam
adopts a crystalline structure with only a few columns of bubbles (double staircase and
staircase structure). Below a certain channel width the staircase  structure
becomes unstable and a transition towards a bamboo structure is observed
\cite{drenckhan05}.
These structures are sensitive to both the average bubble size and the
polydispersity of the foam, especially in the transition region between staircase and
bamboo, where small increases or decreases in the bubble size can push the structure either way.  
In general, a bamboo foam is obtained when the channel width is such that $a_2 \le 2\, R$.

Tests were also carried out with $b = 2$~cm and $b_1 = b_2 = 0$~cm (see Fig.~\ref{fig:channelheleshaw}),
giving a channel width
of $w = 6$~cm.  These `straight-walled' tests were carried out to determine whether 
the upstream corners introduced by the inclusion of the two side obstacles had any effect 
on the flow behaviour within the channels.  
No difference in the flow
behaviour was found between the `stepped' and the straight-walled tests, and the
two sets of results are presented together in the following.

\begin{figure}[htb]
\centering
\subfigure[]{
\includegraphics[width=9cm]{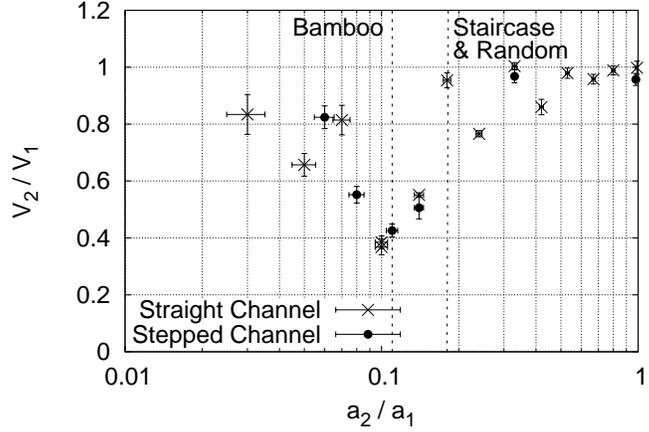}
}
\\
\hspace{0.4cm}
\subfigure[]{
\includegraphics[width=9cm]{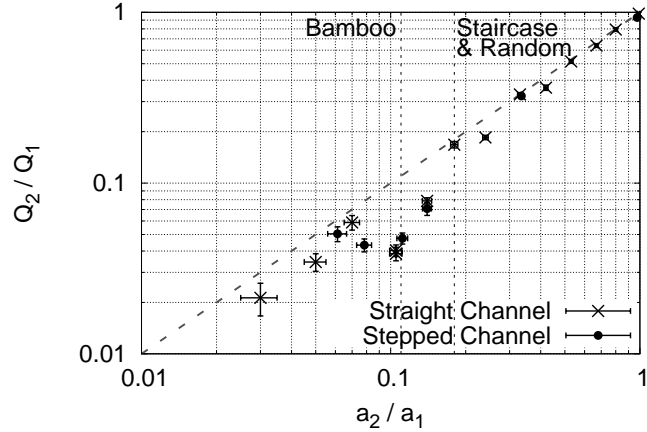}
}
\caption{Dependence of (a) the velocity ratio, $V_2/V_1$,  and  (b) the flux
ratio, $Q_2/Q_1$, on the channel width ratio, $a_2/a_1$, for the enclosed
channels. The bubble size is $R=0.25$ cm and $a_1+a_2= 4$ cm. The dotted vertical
lines
indicate the limits of the transition zone in which a mixture of single staircase and 
bamboo foam was observed in the narrow channel. The regions of the graph in which only bamboo 
or where only staircase or random foams were observed are labeled.}
\label{fig:encvelfluxratio}
\end{figure}

In Fig.~\ref{fig:encvelfluxratio}(a) we plot the ratio of the average bubble
velocities  in each of 
the two channels as a function of the ratio of channel widths $a_2/a_1$.
If two or more bubbles span the narrow channel there is
little difference between the velocities in both channels ($V_2/V_1 \approx 1$).  However, once
the transition to a bamboo foam occurs, the flow in
the narrow channel becomes much slower than in the larger one. The velocity ratio reaches a
minimum for a narrow channel width of the order of  $a_\text{min}= 1.5\, R$.
For very small channel widths, the velocity in the narrow channel increases again  (analogously
to the findings of Bertin {\it et al.} \cite{bertin99}), recovering to within $20\%$ of
the velocity in the wide channel.

The corresponding ratio of the fluxes in the two channels is shown as a function
of the channel width ratio in Fig.~\ref{fig:encvelfluxratio}(b). This plot is
obtained by simply multiplying the velocity ratios of
Fig.~\ref{fig:encvelfluxratio}(a) by the factor $a_2/a_1$. Consequently, in the
staircase/random regime the flux ratio is equal to the channel width ratio, but
once the transition to bamboo structure occurs there is a significant drop in
the flux in the narrow channel.  As the channel width ratio is reduced from
$0.1$ to $0.07$ the flux ratio then increases again. It is significant that when the
foam has a bamboo structure there is a interval in which reducing the size of the
narrow channel causes an {\emph{increase}} in the flux through the narrow
channel.

\subsection{Open-Ended Channels}

In an open-ended channel test, the obstacles were positioned at the end of the Hele-Shaw cell (with plate
separation, $h = 0.2$~cm) so that the foam passing through the two channels
vented to the open air.  Tests were carried out for $a_2$ ranging from $2$~cm
down to $0.1$~cm, and for three different total fluxes in the system, $100$, $50$ or $25$~cm$^3$/min. In these experiments $A$ was in the range $0.25$ to $0.29~{\rm cm}^2$, with an equivalent diameter of $2R\approx0.6$~cm, and with a maximum polydispersity index of $25\%$.

\begin{figure}[htb]
\centering
\subfigure[]{
\includegraphics[width=8cm]{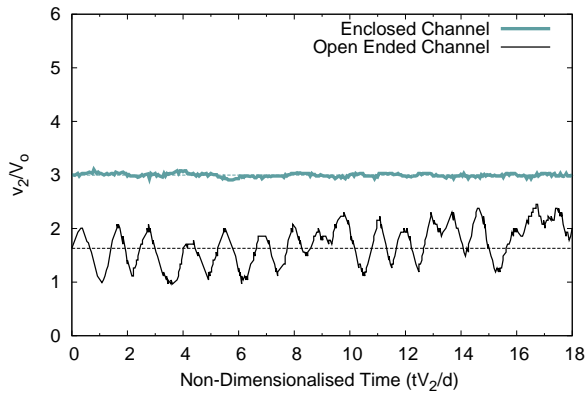}
}
\hspace{0.4cm}
\subfigure[]{
\includegraphics[width=8cm]{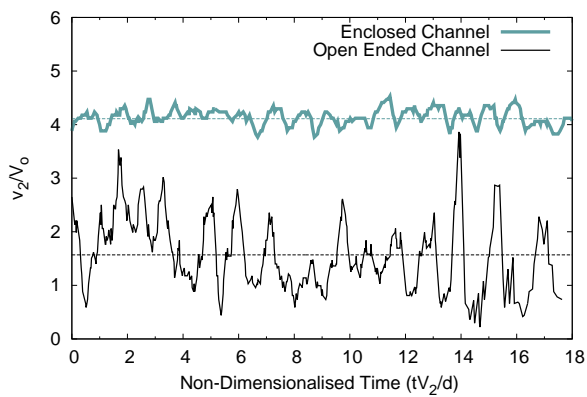}
}
\caption{Velocity, non-dimensionalised by the reference velocity $V_0$, versus dimensionless time 
for both the enclosed and open-ended channels. (a) For a channel of width $0.38$~cm and
(b) for a channel of width $0.12$~cm. 
Large fluctuations are observed for open-ended channels.  
The mean value $V_2$ for each case is indicated with a dashed line.
The velocity in the enclosed-channel data is offset by +2 to improve clarity.}
\label{fig:velocitytimegraphs}
\end{figure}

Similar results to those for the enclosed channels were obtained when staircase or random
foam structures were observed in both channels.
However, once the transition to bamboo foam had occurred in the narrow channel an
important end effect was observed, with the introduction of significant velocity
oscillations in the narrow channel at lower values of $a_2$. 
In this case, all the films of the bamboo structure move at the same velocity $v_2(t)$, that varies with time around its average value $V_2$. 
Fig.~\ref{fig:velocitytimegraphs} shows  $v_2(t)$, divided by the reference  velocity $V_0$ (upstream of the channels), as a function of time, for two different bamboo structures.
For the enclosed tests there is very
little, if any, variation of the velocity with time.  
For the open ended tests however, there are significant variations in the
velocity with time, with the amplitude of these oscillations increasing as the channel width decreases. In Fig.~\ref{fig:velocitytimegraphs}, the time is rescaled by 
$d/V_2$, with $d$ the average distance between two lamellae in the bamboo structure. A lamella reaches the end of channel 2  every $t=d/V_2$. This period coincides with the period of velocity oscillations, which are due to film pinning  at the exit. 

\begin{figure}[htb]
\centering
\includegraphics[width=10cm]{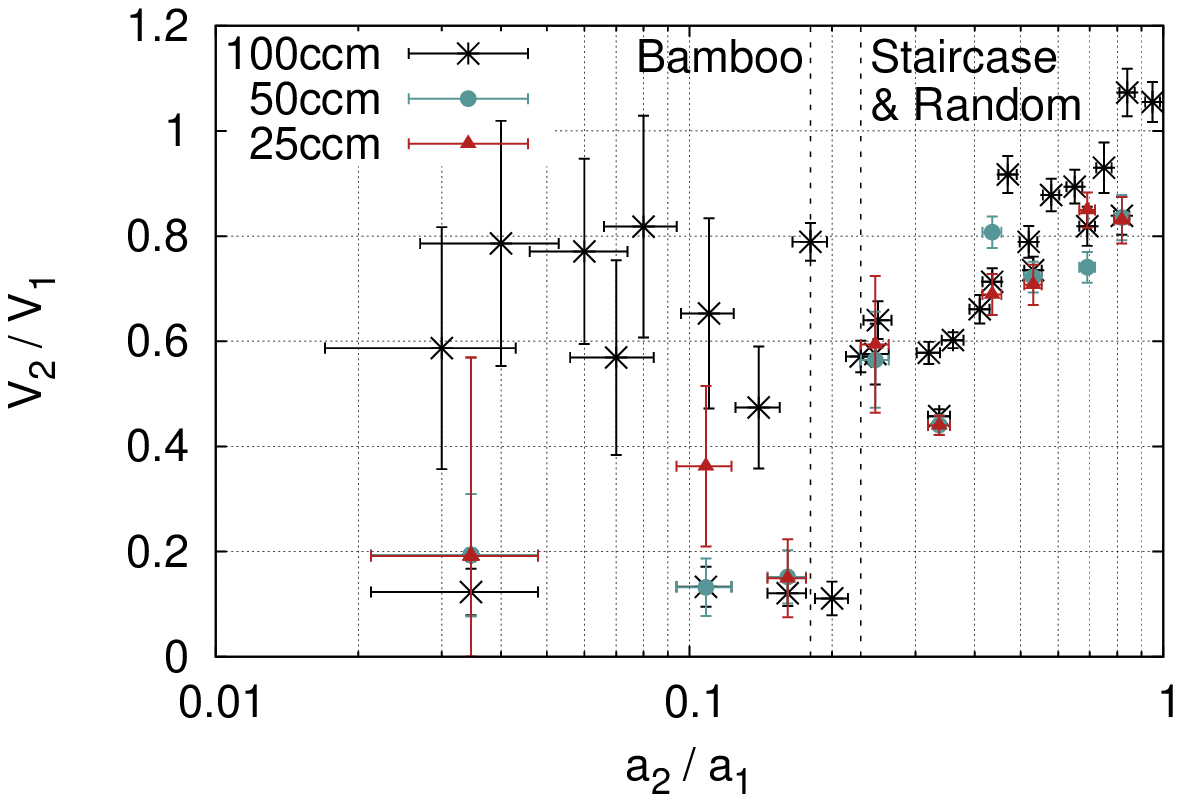}
\caption{
Velocity ratio, $V_2/V_1$, plotted as a function of the channel width ratio, for open-ended channels.}
\label{fig:openvelfluxratio}
\end{figure}

The ratio of the time-averaged velocities in the two channels is plotted as a function of the
ratio of the channel widths in Fig.~\ref{fig:openvelfluxratio}, which 
should be compared to Fig.~\ref{fig:encvelfluxratio}(a). The large
oscillations in the velocity of a bamboo foam in the narrow channel lead to 
large error bars in this region.  There is  a  wide scatter in the data obtained at different fluxes when the foam has a bamboo structure, but no clear dependency has been evidenced.

\section{Model}
\label{model}

\begin{figure}[htb]
\centering
\includegraphics[width=8cm]{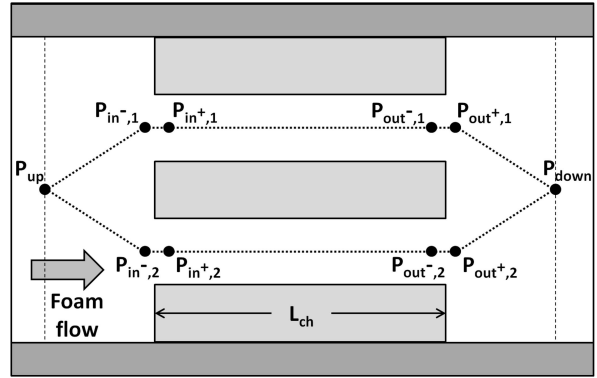}
\caption{\label{fig:schema_pression}Schematic diagram of the experiment
showing the points at which the pressures $P_\text{up}$,
$P_{\text{in}^-}$, $P_{\text{in}^+}$, $P_{\text{out}^+}$, $P_{\text{out}^-}$ and
$P_\text{down}$ are defined.}
\end{figure}

The pressure distribution in the experiment is shown schematically in 
Fig.~\ref{fig:schema_pression}. We define two reference pressures: $P_\text{up}$, the
pressure upstream of the channels, where the pressure distribution perpendicular to the direction of
flow is still homogeneous; and $P_\text{down}$, which for the open-ended channels is 
atmospheric pressure, and  for the enclosed channels is the pressure 
downstream of the channels where the flow becomes uniform again.
The pressure drop $P_\text{up}-P_\text{down}$ can be decomposed into five contributions: (i)
$P_\text{up}- P_{\text{in}^-}$, due to both the viscous friction on the plates before
entering the channel, and to the elasto-plastic stress in the foam, arising from the strong 
deformations that occur in the foam in this region; (ii)  $P_{\text{in}^-}-
P_{\text{in}^+}$ due to a possible high film curvature at the channel entrance, inducing a
Laplace pressure drop (as the bubble size may be of the order of the channel
size this contribution can not be described in the frame of a
continuum model and requires separate treatment); (iii) $P_{\text{in}^+}-
P_{\text{out}^-}$ due only to the viscous force exerted by the channel walls, 
since we have confirmed that the flow is plug-like in the two channels; (iv)  $P_{\text{out}^-}-
P_{\text{out}^+}$, the equivalent of (ii) for the exit; and 
finally, (v) $P_{\text{out}^+}- P_\text{down}$, the equivalent of (i) for the exit. 

The aim of this section is to explain these different contributions, allowing us to 
interpret the experimental velocity measurements in \textsection \ref{comp}.

\subsection{Viscous forces}
\label{sec:pressure_viscous}

The pressure drop associated with the flow of a  bamboo foam
moving at 
velocity $ \mathbf{v}= V \mathbf{e}_x$ in a channel  oriented in the direction
$\mathbf{e}_x$ is directly related to the total length of meniscus in contact
with the wall. In this geometry, the viscous force exerted by the wall is indeed
localised along the meniscus and it can be expressed, per unit length of
meniscus, as
\begin{equation}
\mathbf{f}_\text{film} = - \lambda \gamma \, Ca^{2/3} \mathbf{e}_x,
 \label{vectorf}
\end{equation}
 where $\lambda$ is a dimensionless prefactor, $\gamma$ is the surface tension
and
$Ca$ the capillary number, defined as $Ca=\eta V/\gamma$, $\eta$ being the
dynamic viscosity of the foaming solution.

\begin{figure}[htb]
\centering
{
\includegraphics[width=5cm]{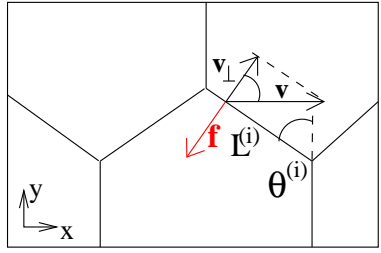}
}
\caption{A meniscus of length $L^{(i)}$ and orientation $\theta^{(i)}$, travelling
at velocity $\mathbf{v}$.  
The normal velocity, $\mathbf{v}_{\perp}=(\mathbf{v}\cdot \mathbf{n})\mathbf{n} $, and the
viscous force acting on the film, $\mathbf{f}$, are indicated.}
\label{fig:Lproj}
\end{figure}

However, for a more general 2D foam structure, the menisci can have an
arbitrary orientation $\theta$ with respect to the channel orientation (see Fig.~\ref{fig:Lproj}). 
The viscous force per unit length of meniscus decreases as
$\theta$ increases. Rather than the total length $L$ of meniscus, 
it was shown by Cantat et al.\cite{cantat04} that the projected length perpendicular to the flow determines the total
viscous force exerted on a moving film, in agreement with their experimental observations.
Similarly, a general vectorial relation for the viscous force was proposed in
\cite{kern04}:
\begin{equation}
\mathbf{f} = f(\mathbf{v}\cdot\mathbf{n}) \mathbf{n} \, ,
\label{fvect}
\end{equation}
where $\mathbf{n}$ is the normal to the film. $f$ 
is the scalar viscous force,
obtained from eq.~\eqref{vectorf}, and is defined by:
\begin{equation}
f(\mathbf{v}\cdot\mathbf{n}) = - \lambda \gamma \, (\eta \mathbf{v}\cdot\mathbf{n}/\gamma)^{2/3}  \text{~ .}
 \label{scalarf}
\end{equation}

For a 2D foam moving without deformation at a uniform velocity $\mathbf{v}=V \mathbf{e}_x$
in a channel of cross-sectional area $S= ah$, 
the viscous pressure drop $\Delta P_\text{visc}$ can be
calculated from the force balance on the foam, with inertia negligible in our
parameter range.   Using eq.~\eqref{fvect} projected along the $\mathbf{e}_x$
direction and summing the contributions from all the menisci in contact with the
 bounding plates, we obtain
\begin{equation}
\Delta P_\text{visc} =  \frac{\gamma \lambda}{S} \left({\frac{\eta
V}{\gamma}}\right)^{2/3}  \; \sum_i \left (L^{(i)} \,\cos^{5/3}\theta^{(i)} \right )
=
 \frac{\gamma \lambda}{S}\: Ca^{2/3}  \; \sum_i \left (L^{(i)} \,\cos^{5/3}\theta^{(i)} \right )
\text{~ ,}
 \label{deltap}
\end{equation}
where $L^{(i)}$ and $\theta^{(i)}$ are the length of the meniscus denoted by index $i$, and its
orientation with respect to the channel axis respectively.

 The experiments in \cite{cantat04} gave a value of $\lambda \approx 38 \pm
10$\% for both bamboo and single staircase structures in a similar geometry.
Raufaste {\it et al.} \cite{raufaste09} extended this work to a more
extensive 2D random foam, 
and also considered the effect of liquid fraction.
Their measurements yielded the empirical relation:
 \begin{equation}
\lambda = (10.27\pm 0.52) \left({\frac{R_p}{\sqrt{A}}}\right)
^{-0.48 \pm 0.02} \text{~ ,}
\label{lambda}
 \end{equation}
  \noindent where $R_p$ is the radius of curvature of the
meniscus at the wall, and $A$ is the bubble area as seen from
above. This relation holds for $R_p/\sqrt{A}$ between 0.01 and 0.35, corresponding to liquid fractions between 0.01\% to 30\%.
The ratio $R_p^2/A$ can be calculated from the liquid fraction  $\phi$ of the foam using\cite{raufaste09}
\begin{equation*}
\frac{R_p^2}{A} = \frac{\phi} {{2C_v + C_{hw}\frac{\mathcal P}{h}}}    \label{rpsqrta}\text{~ ,}
\end{equation*}
\noindent where $h$ is the plate
separation in the Hele-Shaw cell and $\mathcal P$ is the bubble perimeter. $C_v$
and $C_{hw}$ are constants associated with the shapes of the vertical and 
horizontal Plateau borders respectively; for a purely hexagonal foam they are\cite{raufaste09}
 $C_v = \sqrt{3}-\pi/2$ and $C_{hw} = 2 - \pi/2$. 

However, for a foam in a narrow channel the structure is generally not
hexagonal: there are pentagonal bubbles at the walls, and in a very 
narrow channel the bubbles are rectangular (bamboo). $C_{hw}$ is independent of the individual
bubble shape, but $C_v$ varies with changes to the bubble geometry. 
Following \cite{raufaste09}, it has been derived by us to be
$1 + \sqrt{3} /2 - \pi /2$ for 
regular pentagonal bubbles touching the cell side and $2 - \pi/2$ for
bamboo bubbles (touching both cell sides).

Using the relevant parameters for each foam structure, depending on the liquid
fraction and the bubble shape, we obtain values of $\lambda$ for our experiments
in the range 34 to 58.

\subsection{Pressure drop for the bamboo structure}

For the bamboo structure, eq.~\eqref{deltap} can be rewritten explicitly as a
function of the bubble volume $\Omega$. The length of meniscus touching the wall
is, per bubble, $L=2(a+h)$, with an angle $\theta=0$; $a$ and $h$ are
the width  and height of the channel respectively. The distance between
two films is $d= \Omega/(ah)$ and so the number of films in a channel of length
$L_\text{ch}$ is $L_\text{ch}/d$. We therefore 
predict the pressure gradient along a channel to be

\begin{equation}
\frac{\Delta P_\text{visc,bam}}{L_\text{ch}} = \frac{\eta V \lambda}
{Ca^{1/3}} \: \frac{2 (a+h)}{\Omega}  \text{~ .}
 \label{deltap_bamboo}
\end{equation}

Kovscek et al \cite{kovscek03} consider the case $a=h$ and
the apparent viscosity of the foam $\eta_f$  is determined
as defined by the
relation
\begin{equation}
\label{eq:Darcy}
 \frac{\Delta P_\text{visc}}{L_\text{ch}} = \frac{\eta_f}{K} \: V_d  \text{~,}
\end{equation}
where $K$ is the   absolute  permeability of the medium (as defined for a Newtonian fluid),
and $V_d$ is the Darcy (or filtration, or superficial) velocity, which in our case is the mean cross-sectional velocity: $V_d=\psi V$, $\psi= (a_1+a_2)/w$ being the porosity. 
 The permeability $K$ scales as $a^2$, i.e. $K = \beta a^2$, with
$\beta$ a constant depending on the geometry. Note that $\beta$ could differ considerably from unity if
the aspect ratio of the pore cross section itself is very
different from unity. 
 Using these notations, relation
\eqref{deltap_bamboo} can thus be transformed into
\begin{equation}
\frac{\Delta P_\text{visc,bam}}{L_\text{ch}} =  \frac{\eta V_d}{\psi K} \:
\frac{\lambda}{{Ca}^{1/3}}\: \frac{4 K^{3/2}}{\Omega \beta^{1/2}}  \text{~ ,}
 \label{deltap_bamboo_2}
\end{equation}
which by comparison with eq.~\eqref{eq:Darcy} shows that the effective viscosity of
the foam scales with $K^{3/2}$:
\begin{equation}
 \eta_f = \eta \: \frac{\lambda}{{Ca}^{1/3}}\: \frac{4 K^{3/2}}{\Omega \psi \beta^{1/2}}  \text{~ .}
\end{equation}

Hence the improved foam penetration observed in the low permeability domains in the experiments
of Bertin et al. \cite{bertin99} results from this $K^{3/2}$ dependence.  We also note
that increasing the overall flow rate (i.e. increasing $Ca$)  improves the efficiency of the displacement, as
measured in \cite{guillen12}.

\subsection{Pressure drop for a staircase structure and a randomly-oriented
hexagonal 2D foam}

Eq.~\eqref{deltap_bamboo}, however, does not predict the significant velocity
decrease at the staircase to bamboo transition, when the amount of meniscus
in contact with the wall suddenly increases, making the  sum in eq.~\eqref{deltap} much larger at given bubble volume and
channel width \cite{cantat04,raven06b}.

\begin{figure}[htb]
\centering
\includegraphics[width=6cm]{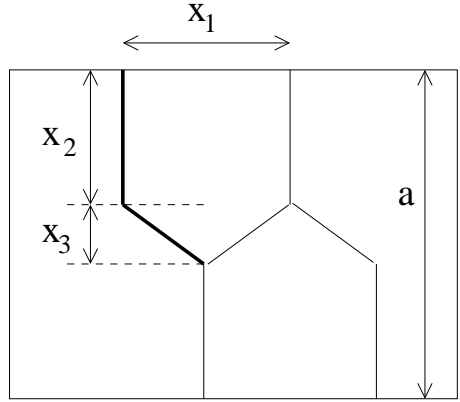}
\caption{Sketch of a staircase structure.  }
\label{staircase}
\end{figure}

The different edge lengths and orientation of the staircase can be determined
from the bubble volume $\Omega$ and the channel width $a$ and thickness $h$.
Three characteristic lengths are depicted in Fig.~\ref{staircase}: $x_1$, $x_2$
and $x_3$; they are related to each other through the relations $2x_2+ x_3= a$ and
$x_1 (x_2 + x_3/2)=  \Omega/h$, whence $x_1=(2\,\Omega)/(ah)$.
Neglecting the deformation of the bubble induced by viscous forces
\cite{green09}, we assume that the angles between the edges are $2 \pi/3$. We
can then deduce $x_1 = 2 \sqrt{3}x_3$, which immediately leads to $x_3= \Omega
/(\sqrt{3} a h)$. 

A length $L_\text{ch}$ of foam consists of $N = h L_\text{ch} a /\Omega$ bubbles. The meniscus network in contact with the plates has $N$ vertical segments of length $h$, $2N$ segments of length $x_2$ and $2N$ segments of length $x_3$, at an angle $\theta= \pi/3$  (see Fig.~\ref{staircase}). 
Finally, eq.~\eqref{deltap}  leads to:

\begin{equation}
\frac{\Delta P_\text{visc,stair}}{L_\text{ch}} = \frac{\eta V \lambda}{Ca^{1/3}} \: \left( \frac{a+h}{\Omega}+ \frac{2^{1/3}-1}{3^{1/2}S}
\right)
\text{~ .}
 \label{deltap_stai}
\end{equation}

The pressure drop induced by the motion of a randomly-oriented
hexagonal 2D foam is obtained by averaging the contribution of regular hexagons  over all possible
 orientations. We assumed $a\gg h$ and $a\gg \sqrt(\Omega/h)$ so that the correction arising from the lateral boundaries become negligible.

\begin{equation}
\frac{\Delta P_\text{visc,rand}}{L_\text{ch}} = 1.99\: \frac{\eta V \lambda}{Ca^{1/3}} \: \left( \Omega h \right)^{-1/2}  \text{~ .}
 \label{deltap_rand}
\end{equation}

\subsection{Capillary pressure}
\label{model_cap}

\begin{figure}[htb]
\centering
\includegraphics[width=8cm]{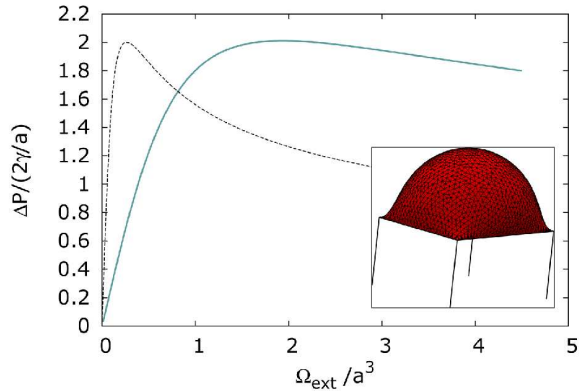}
\caption{Surface Evolver  \cite{brakke92} simulation of a bamboo film emerging from a channel with
a rectangular cross-section $a\times h$ with $h=0.8\,a$. 
Main graph:  Ratio of the capillary pressure  across the film to $2 \gamma/a$,
as a function of the bubble volume external to the channel $\Omega_\text{ext}$ divided by $a^3$. For comparison, we show the result obtained for a cylindrical tube of radius 
$a/2$ (dashed line).
Inset: Image of the emergent film.  The film is pinned all around the exit from the channel.
}
\label{fig:pinnedbubble}
\end{figure}

We next determine the additional pressure drop induced by a bubble
exiting the channel.
When a film emerges from a channel, we have observed that it
remains pinned to the exit corners, and is progressively deformed by the
constant gas flux, until it eventually breaks or depins.
We consider the capillary pressure across the film, $\Delta P_\text{cap}$,
to be given by the Young-Laplace equation:
\begin{equation}
\Delta P_\text{cap} = 2 \gamma  \left({\frac{1}{r_1}}+{\frac{1}{r_2}}\right)
\text{~ ,}  \label{pcapillary}
\end{equation}
where $r_1$ and $r_2$ are the two radii of curvature of this bubble shape.
As an illustrative example, we computed the film shape with Surface Evolver \cite{brakke92} for the
case of a bamboo film exiting a rectangular channel, with an arbitrary value of $h=0.8\,
a$, pinned on the channel boundary (Fig.~\ref{fig:pinnedbubble}). The pressure 
reaches its maximal value $\Delta P_\text{cap}^M \sim 2 \gamma (1/a+1/h) $
for an external volume $V \sim a^3$ before decreasing slowly as the external
volume increases.  This maximal pressure has to be overcome before the flow in the
channel can proceed. After rupture or depinning this film is replaced by another film and,
after a short transient, the pressure reaches a value close to its maximal value again.

In general, the capillary pressure  at the exit from the channel is
\begin{equation}
 \Delta P_\text{cap}= P_{\text{out}^-}- P_{\text{out}^+}
 = 2 \gamma \left( \left\langle \frac{1}{r_1} \right\rangle + \left\langle \frac{1}{r_2} \right\rangle \right) \, ,
\label{eq_pcap}
\end{equation}
where $\langle 1/r_1 \rangle$ is a film's time-averaged curvature
in the horizontal plane and
 $\langle 1/r_2 \rangle$ its time-averaged  curvature in the vertical
plane parallel to the channel's side walls.

Instead of doing extensive numerical simulations, and in order to obtain analytical predictions, we choose to use an approximated but simple value for this capillary pressure drop.  
For the enclosed channel experiments, the top and bottom plates are smooth and the film is
free to slide on them. We thus  use $\langle 1/r_2 \rangle = 1/h$ for the open-ended 
channel and $\langle 1/r_2 \rangle = 0$ for the enclosed structure. For
the bamboo structure  $\langle 1/r_1 \rangle \approx 1/a$ for both open-ended and enclosed cases. 
For the other structures, it is more difficult to give a precise analysis of
the bubble  shape at the exit. A good order of magnitude is obtained with
$\langle 1/r_1 \rangle \approx 1/R$ for the staircase structure ($R$ being the typical bubble radius, see \textsection \ref{setup}) and  $\langle
1/r_1 \rangle \approx 0$ for a random foam, as in this situation $a\gg h$ and pinning on the side walls is expected to have a negligible influence.

The capillary pressure at the entrance $P_{in,+}-P_{in,-}$ is more difficult to predict and is discussed in section \ref{comp_enclosed}.

\subsection{Pressure drop upstream of the entrance}

The foam velocity  is very inhomogeneous close to the channel entrance, with
both stagnation points and high-velocity domains.  The foam is thus strongly
sheared and internal elastic and viscous stresses govern the pressure
fields in addition to the friction on the plates discussed in \textsection
\ref{sec:pressure_viscous}. As the velocity field in front of each channel is
different, $P_\text{up}- P_{\text{in}^-,1}$ could {\it a priori} significantly differ
from $P_\text{up}- P_{\text{in}^-,2}$.
To test this assumption, we carried out experiments with two channels of equal
width in an off-centre position in the Hele-Shaw cell ($a_1=a_2$,
$b_1=0$~cm, $b_2=4$~cm), in a open-ended geometry. We obtained the same velocity
in both channels within $10\%$. This  implies that  $P_{\text{in}^-,1}- P_\text{down}$ and
$P_{\text{in}^-,2}- P_\text{down}$ are close, {\it i.e.}  $(P_{\text{in}^-,1}-
P_{\text{in}^-,2})/(P_{\text{in}^-,1}- P_\text{down}) \ll 1$. This indicates that even a very
asymmetrical entrance leads to a negligible difference between $P_{\text{in}^-,1}$ and
$P_{\text{in}^-,2}$, and we will consider both values as equal in the following.

\section{Comparison of the model with the experimental measurements }
\label{comp}

\subsection{Enclosed channel}
\label{comp_enclosed}

In order to predict the velocity ratio between the two channels, we take into
account both the viscous and the capillary pressure at the exit, discussed in \textsection
\ref{sec:pressure_viscous} and \ref{model_cap}, with the assumption that  $\Delta
P_\text{visc,1} + \Delta P_\text{cap,1} = \Delta P_\text{visc,2} + \Delta P_\text{cap,2}$. Using
eqs.~\eqref{deltap} and \eqref{eq_pcap}, we deduce the velocity ratio as a
function of the physico-chemical parameters, the experimental foam structure and
the velocity in the wide channel:

\begin{equation}
\frac{V_2}{V_1}=\frac{1}{S_2^{3/2}}\left[ S_1 -\frac{2}{Ca_1^{2/3}r_2}
\right]^{3/2} \text{~ ,}
\label{eq_vtheo_cor}
\end{equation}
\noindent with
\begin{equation}
S_j= \frac{\lambda}{a_j\,h} \; \sum_i \left (L_j^{(i)} \,\cos^{5/3}\theta_j^{(i)} \right ) \text{~ , ~}j \in \{1,2\}
\text{~ .}
\label{eq_K}
\end{equation}

To derive this expression, we have used two facts: that the terms containing $r_1$ cancel out since the two channels have the same height, and that $\langle
1/r_2 \rangle \approx 0$ in the largest channel since the foam is always random there. The relative importance of the viscous and capillary terms depends on the velocity, or equivalently on the capillary number in the wide channel $Ca_1$,
and hence depends on the total flux. It also depends on the channel length,
through the summation over all menisci in contact with the walls.

\begin{figure}[htb]
\centering
\subfigure[]{
\includegraphics[width=8cm]{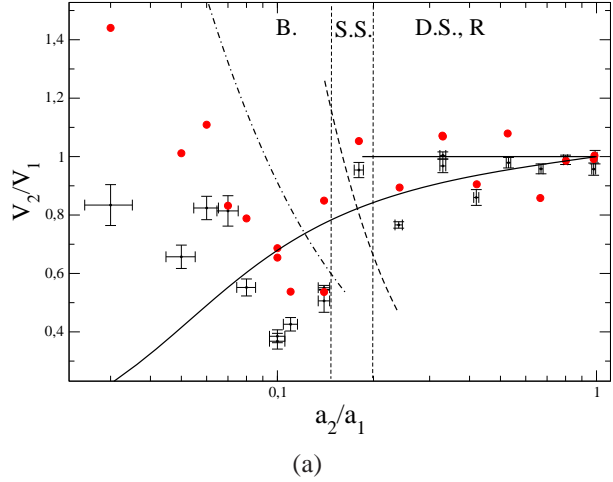}
}
\\
\vspace*{0.8cm}
\subfigure[]{
\includegraphics[width=8cm]{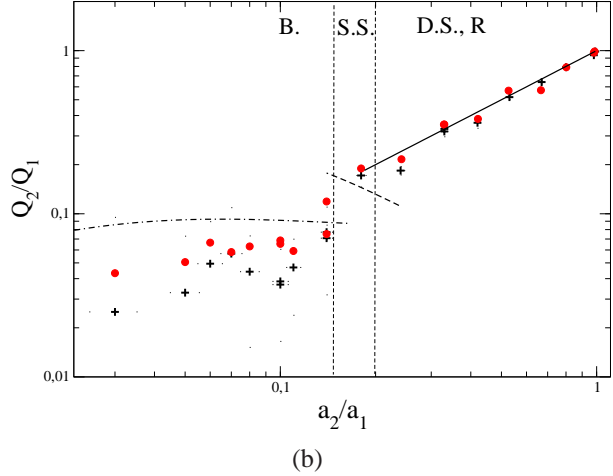}
}
\caption{(a) Comparison of the experimental data of Fig.~\ref{fig:encvelfluxratio} with theoretical predictions.  Analytical prediction: horizontal solid line - random foam  structure (R) in the narrow channel;  dashed line - staircase structure (S.S.), eq.~\eqref{repV_stair}; dotted-dashed line - bamboo structure (B. S.), eq.~\eqref{repV_bamb} ($A =0.19$ cm$^2$, $h=0.1$cm and $a_2=4x/(1+x)$ in cm, with $x=a_2/a_1$). 
The functions are plotted in the domains of existence of the corresponding structure, as observed experimentally. The vertical dashed lines that separate these domains are not strict frontiers, and some fluctuations are observed. In the ``random domain'', adjacent to the staircase domain, one data point corresponds to a double staircase structure (D.S.), for which a special treatment would lead to a better theoretical prediction.  
The prediction of eq.~\eqref{eq_vtheo_cor}, determined from the experimental values of the foam
structure and the velocity in the wide channel, is given as {\color{red}$\bullet$}. 
The curved solid line is the velocity ratio expected for a Newtonian fluid \cite{bruus}. The analytical expression is given in Annex.
(b) Same data for the fluxes (without the result for  a Newtonian fluid). 
}
\label{fig:openvelratio_model}
\end{figure}

This prediction is plotted in Fig.~\ref{fig:openvelratio_model} (dots, red online) and
shows a reasonable  agreement with most experimental values.
However,  for the smallest channels, the prediction  overestimates the velocity in the narrow channel. One reason for this may be an overestimation of $\lambda$, as the value given in eq.~\eqref{lambda} was measured with a different foaming solution \cite{raufaste09}.

\begin{figure}[htb]
\centering
\subfigure[]{
\includegraphics[width=3cm]{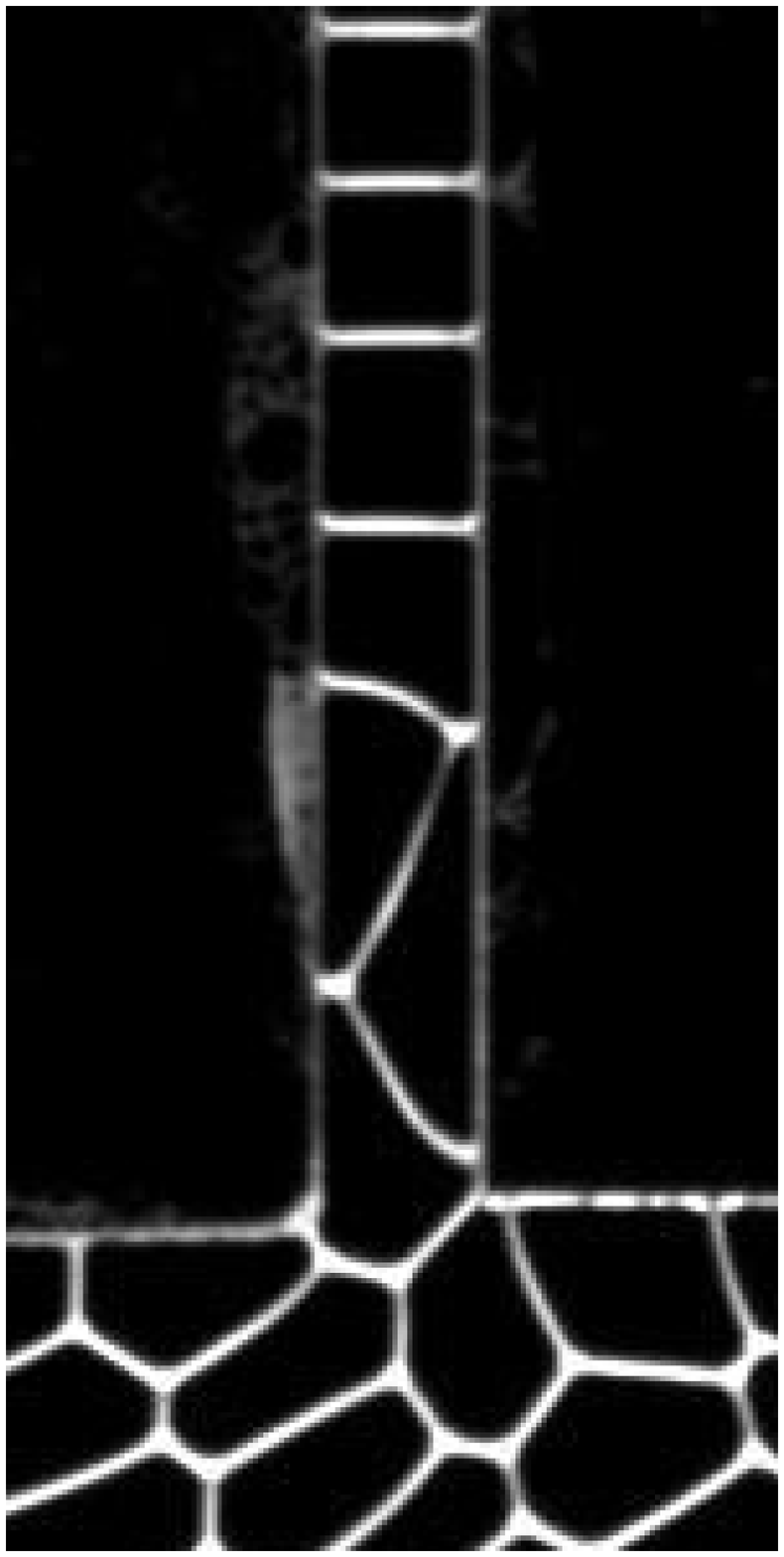}
}
\hspace{0.5cm}
\subfigure[]{
\includegraphics[width=3cm]{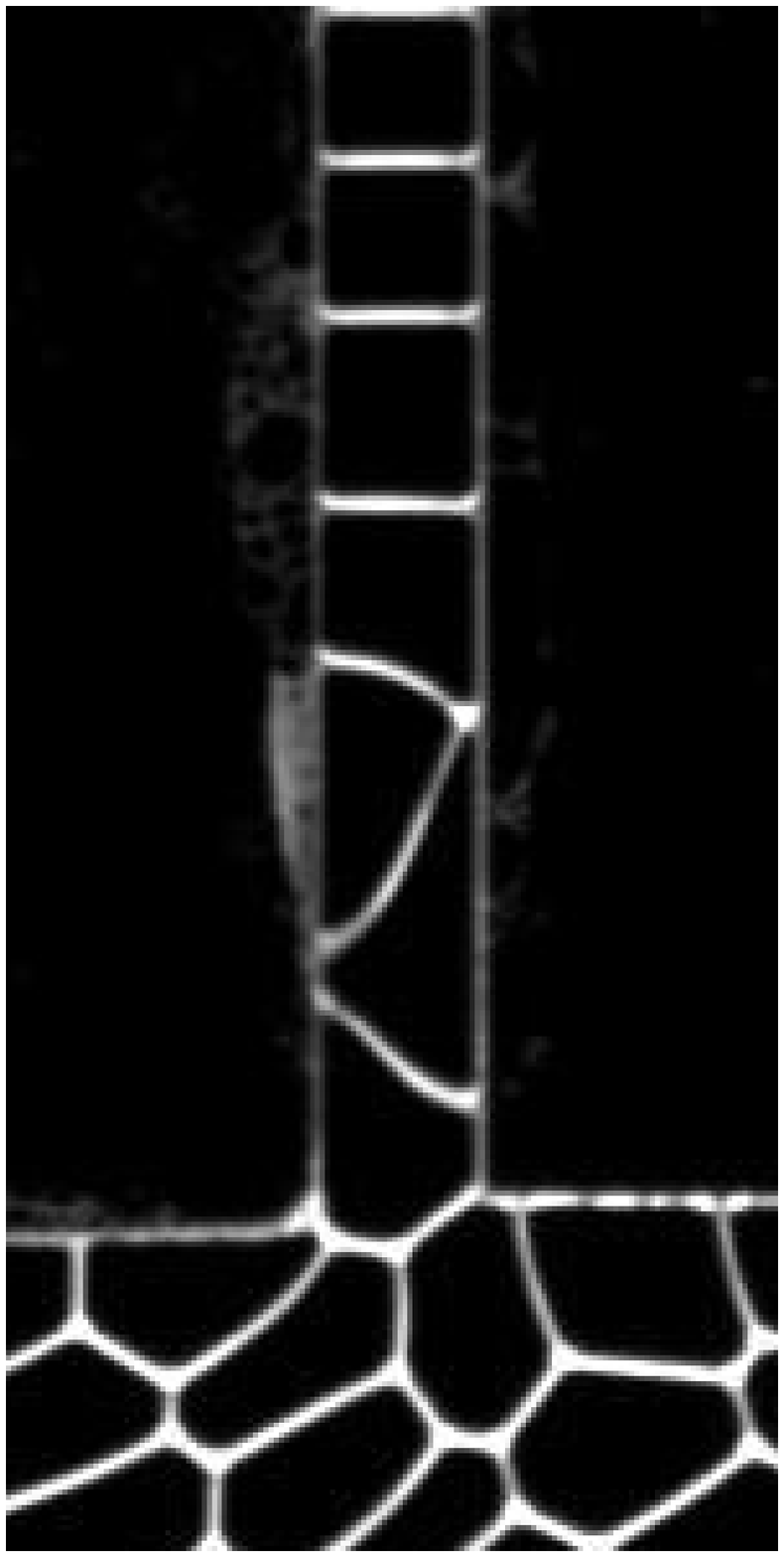}
}
\hspace{0.5cm}
\subfigure[]{
\includegraphics[width=3cm]{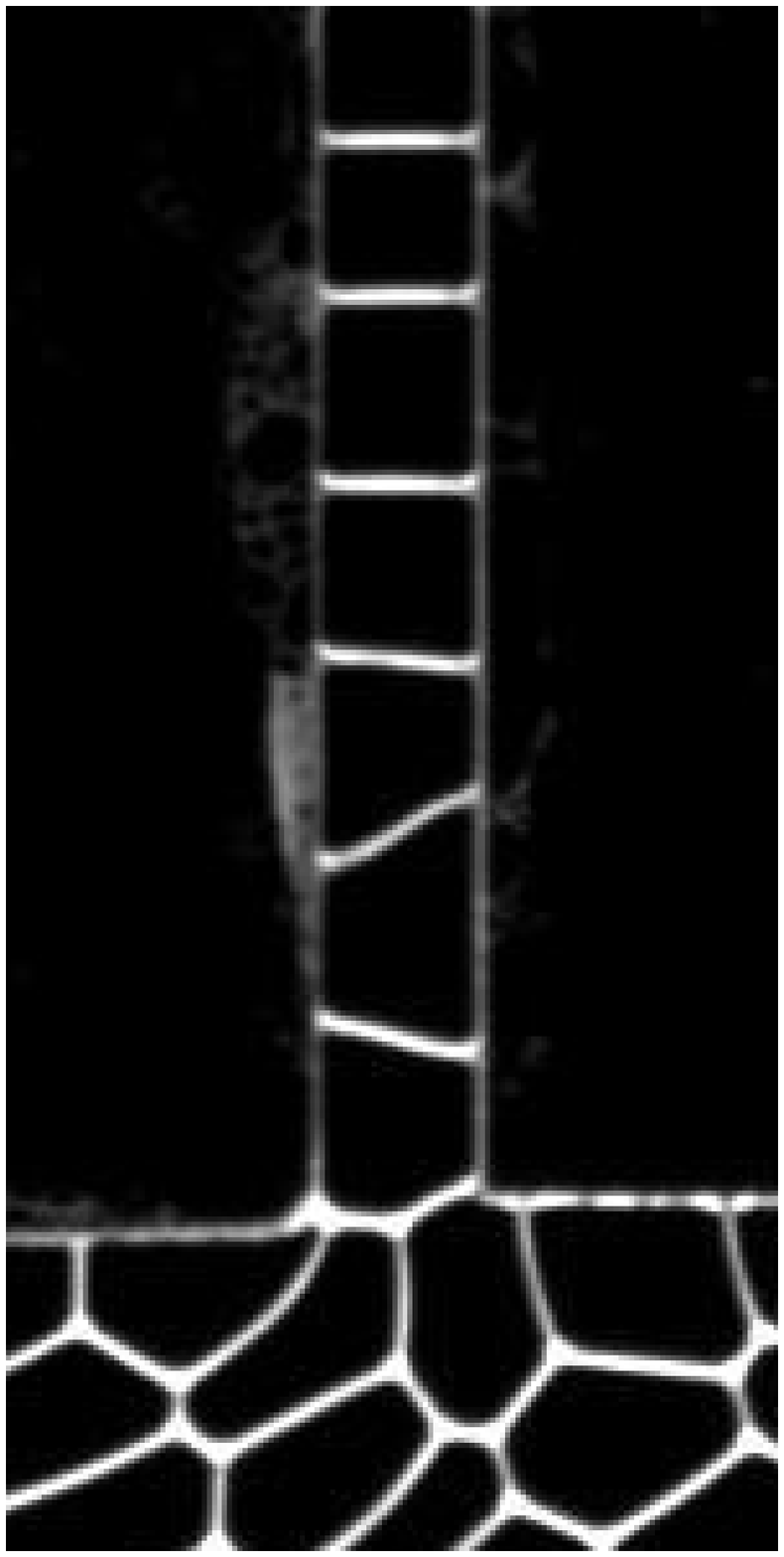}
}
\hspace{0.5cm}
\subfigure[]{
\includegraphics[width=3cm]{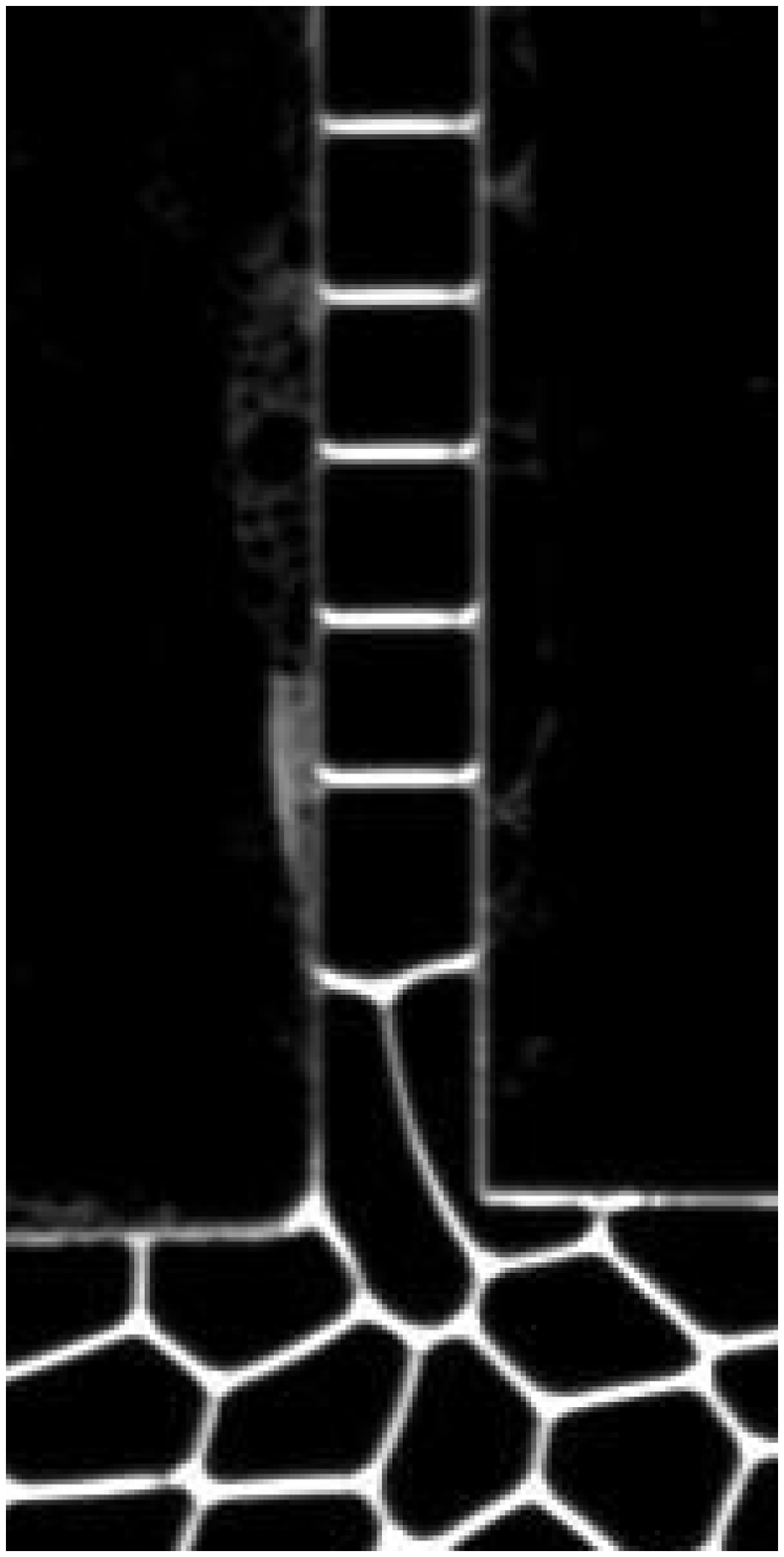}
}
\caption{Images of the foam entering a narrow channel ($a_2=0.49$~cm) to form a
wide bamboo structure.  The average bubble area is $0.25 \pm 0.05$~cm$^2$,
with an equivalent diameter of $0.51 \pm 0.06$~cm.  Images (a)-(c) show three
consecutive frames ($\Delta t = 0.05$~s) of an experiment in which a transient
staircase structure is formed at the entrance to the channel before making the transition to a
wide bamboo structure.  Image (d) shows a single film being stretched
down the centre of the channel -- this phenomena only occurs when the foam forms
a wide bamboo structure.}
\label{fig:entranceeffect}
\end{figure}

Additionally, as noted above, the behaviour of the foam at the
entrance to the narrower channel can be quite complex, as shown in Fig.~\ref{fig:entranceeffect}, with transient staircase
structures being formed that then undergo a transition to bamboo (Figs.~\ref{fig:entranceeffect}(a)-(c)). These transition occur through topological rearrangements, called T1s, during which the bubble contact network is modified.  We also observed that films attached to bubbles outside the narrow channel
could become greatly extended along the centre of the channel
(Fig.~\ref{fig:entranceeffect} (d)). This stretch persists until a T1 occurs,
and can result in a bubble being split into two parts if the corner of the obstacle
pierces the film prior to the T1. 
It is very difficult to
quantify these effects, but there is potentially an induced correction factor for $P_{in,+}-P_{in,-}$ in the
pressure balance equation for the narrowest channels.

Most notably the velocity ratio varies non-monotonically with the
width ratio, with sudden jumps at the structure transitions, an important feature that can be explained by considering the viscous pressure drop only.  Eqs.~\eqref{deltap_bamboo},
\eqref{deltap_stai} and \eqref{deltap_rand} allow us to derive the following analytical predictions
for the velocity ratio as a function of the channel width ratio for every
possible structure in the narrow channel of width $a$ and thickness $h$ (a random foam is always assumed in the
wider channel):

\begin{equation}
\frac{V_\text{bamb}}{V_\text{rand}}= \left(\frac{0.99 \sqrt{A}}{a_2+h} \right)^{3/2},
\label{repV_bamb}
\end{equation}

\begin{equation}
\frac{V_\text{stair}}{V_\text{rand}}= \left(\frac{1.99}{\sqrt{A}}\right)^{3/2}  \left(
\frac{a_2+h}{A}+ \frac{2^{1/3}-1}{\sqrt{3}a_2} \right)^{-3/2}.
\label{repV_stair}
\end{equation}

\vspace*{0.5cm}

These predictions are plotted in Fig.~\ref{fig:openvelratio_model} together with the experimental data. 
The model captures the non-monotonous
behavior, in particular the velocity jump at each structure transition; this is
strikingly different from the behavior of a Newtonian fluid, which shows a smooth decrease in velocity ratio with decreasing channel width ratio. We overestimate the velocity in the narrowest channels due to the
neglect of the capillary pressure, however this purely viscous model is expected to be valid for very long channels, for which the viscous pressure drop becomes much higher than the capillary pressure at the channel ends. 
The model predicts a large increase of $V_2$  as $a_2/a_1$ becomes small, but above a certain velocity the films break \cite{dollet10}, which limits the accessible velocity range.
It should also be noted that the plateau at larger $a_2/a_1$ ratios (i.~e., when the foam structure
is double staircase or random in both channels) is expected from relation~\eqref{deltap_rand} to
be independent of the channels' widths.

\subsection{Intermittency of the flow triggered by the capillary pressure
drop}\label{sec:intermitt}

For an open-ended channel, the shape of any film external to the
Hele-Shaw cell evolves with time.  As discussed in \textsection \ref{model_cap}, the
Surface Evolver was used to calculate this changing pressure drop as a 
film emerges from the channel, and a graph of the dimensionless pressure drop
against the dimensionless external volume is given in Fig.~\ref{fig:pinnedbubble}.  
This curve has a maximum, and if an
experiment is carried out with a low total flux in the system, a situation can
be reached where the driving pressure in the system is not enough to get past
this maximum, and the flow in the narrow channel is therefore halted.

\begin{figure}[h]
\centering
\centerline{
\includegraphics[width=8cm]{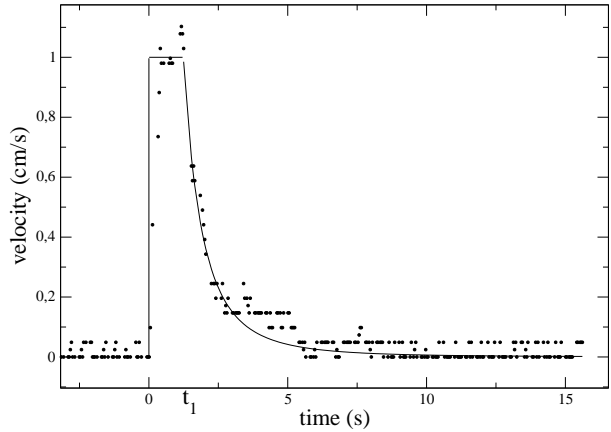}
}
\caption{Velocity of a bamboo film in a narrow channel as a function of time. $\bullet$ Experimental
data. Solid line, for $t>t_1$: prediction of eq.~\eqref{vdet}
$v(t)=v_0\left [ 1 + (t-t_1)/\tau \right ]^{-3}$, with the adjustable parameter $\tau=2s$.  
$v_0$ is the maximal velocity, reached by the lamella 
 between $t=0$, the time at which a pinned bubble
burst, and $t_1=1.24$s, the time at which a new film was pinned at the channel
outlet. }
\label{fig:blockedvelocity}
\end{figure}

This phenomena was observed for $a_2 = 0.13$~cm with a total flux in the
system of $25$~cm$^3$/min.  The foam remained stationary within the narrow channel
until the external bubble burst due to film drainage, resulting in a rapid spike
in velocity before decay back down to zero.  The resulting velocity trace for
this low flux case is given in Fig.~\ref{fig:blockedvelocity}.

The velocity suddenly increases when the bubble bursts (at time $0$ in Fig.~\ref{fig:blockedvelocity})
and reaches a steady value $v_0$ imposed by the pressure drop $\Delta P_0$,
according to
\begin{equation}
 \Delta P_0 = N \, f(v_0) \, \frac{2(a+h)}{a\,h} \text{~ ,}
\end{equation}
where $N$ is the number of bamboo films present in the tube and $f(v_0)$
is the scalar viscous force previously defined in eq.~\eqref{scalarf}.
It should be noted that, as the flux in the wide channel is
much larger than the one in the narrow channel, it can
be considered equal to the imposed total flux, and therefore
$\Delta P_0$ can be considered constant. This means that $v_0$ is
governed by the imposed total flux. Eventually, at a time $t_1$, another film reaches
the end of the tube and becomes pinned.
The film velocity $v(t)$ begins to decrease, as the capillary pressure increases
and a new  blocked situation is reached.

The exact shape of the bubble when the meniscus is pinned at the exit of the channel is not well-controlled and,  in order to get a simple analytical prediction for the intermittent velocity, we assume the simplest possible shape for the pinned contour: a circle of radius $\alpha$. 
We checked that the analogous 2-D calculation for a rectangle of size $h\times 2\alpha$, with $h\gg \alpha$ leads to a very similar prediction, up to a prefactor close to 1. The prediction is also very close to the full numerical calculation for $\Omega_\text{ext}/a^3 \lesssim 1 $ (Fig.\ref{fig:pinnedbubble}). 

As the bubble reaches the exit and remains pinned on the circle of radius $\alpha$, its shape is a spherical cap with radius of curvature $R$.
The external volume is given by
\begin{equation}
\Omega_\text{ext}=\frac{\pi}{2} R(1-\cos \theta_{pin})\left[\frac{R^2}{3}(1-\cos \theta_{pin})^2+ \alpha^2 \right]
\end{equation}
with $\sin \theta_{pin} = \alpha/R$.
In the limit $R \gg \alpha$ we get 
\begin{equation}
\frac{1}{R}= \frac{4\Omega_\text{ext} }{\pi \alpha^4}.
\end{equation}
Since the overall pressure drop is fixed at $\Delta P_0$, the viscous pressure drop
along the channel is simply $\Delta P_0$ less the capillary pressure across the bubble interface:
\begin{equation}
\Delta P_\text{visc}(t)= \Delta P_0 - \frac{4 \gamma}{R} = \Delta P_0 - \frac{16\, \Omega_\text{ext} \,\gamma}{\pi \alpha^4} \text{~ .}
\end{equation}
The external volume of the bubble is also related to the flux:
\begin{equation}
\Omega_\text{ext} = \int_{t_1}^t \pi \alpha^2 \, v(t) dt \text{, }
\label{vol_out}
\end{equation}
which allows us to deduce the viscous pressure  drop as a function of time
\begin{equation}
\Delta P_\text{visc}(t)= \Delta P_0- \frac{16\, \gamma}{\alpha^2} \int_{t_1}^t v(t) dt \text{~ .}
\label{p_inter}
\end{equation}
Using eq.~\eqref{deltap} and assuming that the number of menisci does not change during
the duration of the bubble growth, we have
$v_0^{2/3}/ \Delta P_0 = v(t)^{2/3}/ \Delta P_\text{visc}(t)$, which results in

\begin{equation}
\left(\frac{v(t)}{v_0} \right)^{2/3} = 1 - \frac{16 \gamma }
{\alpha^2\, \Delta P_0}\int_{t_1}^t v(t') dt' \text{~ .}
\label{eq:v_evolution_integral}
\end{equation}
After differentiating with respect to time, eq.~\eqref{eq:v_evolution_integral} becomes a simple differential equation
that can be solved by separation of variables, leading to:
\begin{equation}
\frac{v(t)}{v_0}  = \left[1 + \frac{8 \,\gamma \, v_0(t-t_1)}{\alpha^2\, \Delta P_0} \right]^{-3}= \left[1 + \frac{t-t_1}{\tau}
\right]^{-3}  \text{~ ,}
\label{vdet}
\end{equation}
where the characteristic time for the velocity decay is $\tau = \alpha^2 \Delta P_0 / (8\gamma v_0)$.
As shown in Fig.~\ref{fig:blockedvelocity}, a good agreement is obtained for $\tau = 2$ s. This corresponds to $\alpha= 0.6$ cm.   The corresponding disc area is 1.13 cm$^2$ whereas the actual section of the channel is 0.026 cm$^2$. The discrepancy is larger than expected, for unidentified reasons. The pressure oscillation is thus only  understood qualitatively, and would require a specific study, with a shape  visualisation at the exit, which is beyond the scope of this paper. 

In this situation, the average velocity in the narrow channel strongly depends
on the bubble behavior at the exit (pinning, bursting etc.), which was not controlled 
in the open-ended experiments. This explains the dispersion in the data at low width 
ratios in Fig.~\ref{fig:openvelfluxratio}, in the regime where flow can, notably, be arrested in the narrow channel.

\section{Conclusion}
\label{conclusion}

In this paper, we have measured the velocity of a two-dimensional foam as it
flows through two parallel channels. We have varied the relative width of the
channels at fixed combined width, in order to investigate how changes in the foam
structure inside the narrower channel influence the distribution of fluxes
between the two channels, for a given imposed total volumetric flow rate. Two
geometric configurations were used, one with channels enclosed inside a larger
Hele-Shaw cell, and another one in which the two channels vent to the open air.
They are respectively relevant to (i) the continuous flow of a foam through a
channel within a porous medium, and (ii) to the imbibition of a foam into an
air-filled porous medium or to the flow of a foam along a channel that emerges
into a void much larger than the typical bubble size.

In the bamboo regime the velocity in the narrow channel decreases as the channel width  is increased. This brings a direct experimental evidence 
of the better foam penetration in the smallest pores that had already been observed in real 3D porous media. 
However, when the narrow channel becomes larger than the bubble size, the structure goes from a bamboo structure to a staircase structure. 
This structure transition induces an abrupt increases of the velocity in the narrow channel which dominates over the slow decreases of the velocity with increasing channel width previously discussed. Eventually, when the narrow channel is in the random foam regime (with bubble size much smaller than channel width), the velocity becomes independent of the channel width.
 A theoretical model accounting only for  the viscous dissipation at the contact
between the complex  menisci network and the bounding plates allows us to qualitatively predict the smooth decrease of the velocity in the narrow channel, and the sudden jumps at the structure transitions observed when the narrow channel width increases. Taking into account corrections induced by the capillary pressure across the  films pinned at the channel outlet leads to an almost quantitative velocity prediction, although the theory overestimates to some extent the velocity in the narrower channel.

This work  improves our  understanding of the better foam penetration in the smallest pores and underlines the strong influence of the local organisation 
of the foam. 
For configurations of open-ended channels with a very narrow channel, the pinning effect at the channel exit is much more important and the flow of the
bamboo foam becomes intermittent.  The flow is periodically brought close to a halt while the capillary pressure  across the bubble pinned at the channel outlet builds up, until that bubble bursts or depins. 

Prospects for future work arising from this study include the study of foam dissipation at an intersection of two channels,
and, more generally, the study of foam flow in experimental geometries that resemble the geometry of a random porous medium more closely.

\section*{Appendix}
Calculation of ratio of mean velocities in the Newtonian case. \\
The ratio of the mean velocities for a Newtonian fluid, calculated from a solution of Poiseuille flow \cite{bruus} is  
\begin{equation*} 
\frac{V_2}{V_1}=\frac{ \left[1 - \sum_{n,\text{odd}}^\infty \; {\frac{1}{n^5}} \; {\frac{192}{\pi^5}} \; {\frac{h}{a_2}} \; \tanh\left(n\pi {\frac{a_2}{2h}}\right)  \right]}{ \left[1 - \sum_{n,\text{odd}}^\infty \; {\frac{1}{n^5}} \; {\frac{192}{\pi^5}} \; {\frac{h}{a_1}} \; \tanh\left(n\pi {\frac{a_1}{2h}}\right)  \right]} \text{~ .}
\end{equation*}
This prediction is plotted with the experimental data in figure \ref{fig:openvelratio_model}.  The curve shows a smooth drop off in the velocity as the channel width ratio decreases.

\section*{Acknowledgments}

We are grateful to Alain Faisant, Jean-Charles Potier and Patrick Chasle for
technical support, and to CNRS, the R\'egion Bretagne and the
Institut Universitaire de France for financial support. SJC acknowledges financial support from the project PIAP-GA-2009-251475-HYDROFRAC.


\begin{thebibliography}{0}%
\makeatletter
\providecommand \@ifxundefined [1]{%
 \@ifx{#1\undefined}
}%
\providecommand \@ifnum [1]{%
 \ifnum #1\expandafter \@firstoftwo
 \else \expandafter \@secondoftwo
 \fi
}%
\providecommand \@ifx [1]{%
 \ifx #1\expandafter \@firstoftwo
 \else \expandafter \@secondoftwo
 \fi
}%
\providecommand \natexlab [1]{#1}%
\providecommand \enquote  [1]{``#1''}%
\providecommand \bibnamefont  [1]{#1}%
\providecommand \bibfnamefont [1]{#1}%
\providecommand \citenamefont [1]{#1}%
\providecommand \href@noop [0]{\@secondoftwo}%
\providecommand \href [0]{\begingroup \@sanitize@url \@href}%
\providecommand \@href[1]{\@@startlink{#1}\@@href}%
\providecommand \@@href[1]{\endgroup#1\@@endlink}%
\providecommand \@sanitize@url [0]{\catcode `\\12\catcode `\$12\catcode
  `\&12\catcode `\#12\catcode `\^12\catcode `\_12\catcode `\%12\relax}%
\providecommand \@@startlink[1]{}%
\providecommand \@@endlink[0]{}%
\providecommand \url  [0]{\begingroup\@sanitize@url \@url }%
\providecommand \@url [1]{\endgroup\@href {#1}{\urlprefix }}%
\providecommand \urlprefix  [0]{URL }%
\providecommand \Eprint [0]{\href }%
\providecommand \doibase [0]{http://dx.doi.org/}%
\providecommand \selectlanguage [0]{\@gobble}%
\providecommand \bibinfo  [0]{\@secondoftwo}%
\providecommand \bibfield  [0]{\@secondoftwo}%
\providecommand \translation [1]{[#1]}%
\providecommand \BibitemOpen [0]{}%
\providecommand \bibitemStop [0]{}%
\providecommand \bibitemNoStop [0]{.\EOS\space}%
\providecommand \EOS [0]{\spacefactor3000\relax}%
\providecommand \BibitemShut  [1]{\csname bibitem#1\endcsname}%
\let\auto@bib@innerbib\@empty
\end{thebibliography}%


\begin{thebibliography}{10}

\bibitem{weaire}
D.~Weaire and S.~Hutzler,
\newblock {\em The physics of foams},
\newblock Oxford Univ. Press, Oxford, 2000.

\bibitem{livre_mousse}
I.~Cantat, S.~Cohen-Addad, F.~Elias, F.~Graner, R.~H\"ohler, O.~Pitois,
  F.~Rouyer, and A.~Saint-Jalmes,
\newblock {\em Les mousses. Structure et dynamique},
\newblock Belin, Paris, 2010.

\bibitem{rossen96}
W.~R. Rossen,
\newblock {\em Foam in enhanced oil recovery}, pages 413--465,
\newblock Dekker, Cleveland, 1996.

\bibitem{wang04}
S.~Wang and C.~N. Mulligan,
\newblock An evaluation of surfactant foam technology in remediation of
  contaminated soil,
\newblock Chemosphere {\bf 57}(9), 1079 -- 1089 (2004).

\bibitem{mulligan01}
C.~N. Mulligan, R.~N. Yong, and B.~F. Gibbs,
\newblock Surfactant-enhanced remediation of contaminated soil: a review,
\newblock Engineering Geology {\bf 60}(1-4), 371 -- 380 (2001).

\bibitem{mulligan03}
C.~N. Mulligan and F.~Eftekhari,
\newblock Remediation with surfactant foam of PCP-contaminated soil,
\newblock Engineering Geology {\bf 70}(3-4), 269 -- 279 (2003),
\newblock Third British Geotechnical Society Geoenvironmental Engineering
  Conference.

\bibitem{rothmel98}
R.~Rothmel, R.~Peters, E.~St.~Martin, and M.~DeFlaun,
\newblock Surfactant Foam/Bioaugmentation Technology for In Situ Treatment of
  TCE-DNAPLs,
\newblock Environmental Science and Technology {\bf 32}, 1667--1675 (1998).

\bibitem{hirasaki85}
G.~Hirasaki and J.~B. Lawson,
\newblock Mechanisms of foam flow in porous media : apparent viscosity in
  smooth capillaries,
\newblock Soc. Pet. Eng. J. {\bf 25}, 176--188 (1985).

\bibitem{rossen90a}
W.~R. Rossen,
\newblock Theory of mobilization pressure gradient of flowing foams in porous
  media: I. Incompressible foam,
\newblock J. Colloid Interface Sci. {\bf 136}(1), 1 -- 16 (1990).

\bibitem{kovscek93}
A.~Kovscek and C.~Radke,
\newblock Fundamentals of foam transport in porous media,
\newblock Technical Report, univ. California Berkeley  (1993).

\bibitem{kornev99}
K.~G. Kornev, A.~V. Neimark, and A.~N. Rozhkov,
\newblock Foam in porous media: thermodynamic and hydrodynamic peculiarities,
\newblock Adv. Colloid Interface Sci. {\bf 82}, 127 -- 187 (1999).

\bibitem{bertin99}
H.~Bertin, O.~Apaydin, L.~Gastanier, and A.~Kovscek,
\newblock Foam flow in heterogeneous porous media: Effect of Cross Flow,
\newblock J. Soc. Petrol. Eng. {\bf 4}, 75-- 82 (1999).

\bibitem{kovscek03}
A.~R. Kovscek and H.~J. Bertin,
\newblock Foam Mobility in Heterogeneous Porous Media,
\newblock Transp. Porous. Med. {\bf 52}, 17--35 (2003).

\bibitem{falls88}
A.~H. Falls, G.~J. Hirasaki, T.~W. Patzek, D.~A. Gauglitz, D.~D. Miller, and
  T.~Ratulowski,
\newblock Development of a mechanistic foam simulator: the population balance
  and generation by snap-off,
\newblock SPE reservoir engineering {\bf 3}, 884--892 (1988).

\bibitem{du11}
D.~Du, P.~Zitha, and F.~Vermolen,
\newblock Numerical Analysis of Foam Motion in Porous Media Using a New
  Stochastic Bubble Population Model,
\newblock Transp. Porous. Med. {\bf 86}, 461--474 (2011).

\bibitem{lenormand85}
R.~Lenormand and C.~Zarcone,
\newblock Invasion Percolation in an Etched Network: measurement of a Fractal
  Dimension,
\newblock Phys. Rev. Lett. {\bf 54}(20), 2226--2229 (1985).

\bibitem{lenormand88}
R.~Lenormand, E.~Touboul, and C.~Zarcone,
\newblock Numerical models and experiments on immiscible displacements in
  porous media,
\newblock J. Fluid Mech. {\bf 189}, 165--187 (1988).

\bibitem{maloy85}
K.~J. M{\aa}l{\o}y, J.~Feder, and T.~J{\o}ssang,
\newblock Viscous fingering fractals in porous media,
\newblock Phys. Rev. Lett. {\bf 55}(24), 2688--2691 (1985).

\bibitem{meheust02}
Y.~M\'eheust, G.~L{\o}voll, K.~J. M{\aa}l{\o}y, and J.~Schmittbuhl,
\newblock Interface scaling in a two-dimensional porous medium under combined
  viscous, gravity, and capillary effects,
\newblock Phys. Rev. E {\bf 66}(5), 051603 (2002).

\bibitem{maloy92}
K.~J. M{\aa}l{\o}y, L.~Furuberg, J.~Feder, and T.~J{\o}ssang,
\newblock Dynamics of slow drainage in porous-media,
\newblock Phys. Rev. Lett. {\bf 68}(14), 2161--2164 (1992).

\bibitem{toussaint05}
R.~Toussaint, G.~L{\o}voll, Y.~M{\'{e}}heust, J.~Schmittbuhl, and K.~J.
  M{\aa}l{\o}y,
\newblock Influence of pore-scale disorder on viscous fingering during
  drainage,
\newblock EPL {\bf 71}(4), 583--589 (2005).

\bibitem{toussaint12}
R.~Toussaint, K.~J. M{\aa}l{\o}y, Y.~M{\'{e}}heust, G.~L{\o}voll, M.~Jankov,
  G.~Sch\"afer, and J.~Schmittbuhl,
\newblock Two-Phase Flow: Structure, Upscaling, and Consequences for
  Macroscopic Transport Properties,
\newblock Vadose Zone J. {\bf 11} (2012).

\bibitem{cottin10}
C.~Cottin, H.~Bodiguel, and A.~Colin,
\newblock Drainage in two-dimensional porous media: From capillary fingering to
  viscous flow,
\newblock Phys. Rev. E {\bf 82}, 046315 (2010).

\bibitem{kovscek07}
A.~Kovscek, G.-Q. Tang, and C.~Radke,
\newblock Verification of Roof snap-off as a foam-generation mechanism in
  porous media at steady state,
\newblock Colloids Surf. A {\bf 302}(1-3), 251 -- 260 (2007).

\bibitem{cantat04}
I.~Cantat, N.~Kern, and R.~Delannay,
\newblock Dissipation in foam flowing through narrow channels,
\newblock EPL {\bf 65}, 726--732 (2004).

\bibitem{kern04}
N.~Kern, D.~Weaire, A.~Martin, S.~Hutzler, and S.~J. Cox,
\newblock The two dimensionnal viscous froth model for foam dynamics,
\newblock Phys. Rev. E {\bf 70}, 041411 (2004).

\bibitem{green06}
T.~E. Green, A.~Bramley, L.~Lue, and P.~Grassia,
\newblock Viscous froth lens,
\newblock Phys. Rev. E {\bf 74}(5), 051403 (2006).

\bibitem{grassia08}
P.~Grassia, G.~Montes-Atenas, L.~Lue, and T.~E. Green,
\newblock A foam film propagating in a confined geometry: Analysis via the
  viscous froth model,
\newblock Eur. Phys. J. E {\bf 25}, 39--49 (2008).

\bibitem{raufaste09}
C.~Raufaste, A.~Foulon, and B.~Dollet,
\newblock Dissipation in quasi-two-dimensional flowing foams,
\newblock Phys. Fluids {\bf 21}, 053102 -- 053110 (2009).

\bibitem{dollet10b}
B.~Dollet,
\newblock Local description of the two-dimensional flow of foam through a
  contraction,
\newblock J. Rheol. {\bf 54}, 741 (2010).

\bibitem{dollet07}
B.~Dollet and F.~Graner,
\newblock Two-dimensional flow of foam around a circular obstacle: local
  measurements of elasticity, plasticity and flow,
\newblock J. Fluid Mech. {\bf 585}, 181--211 (2007).

\bibitem{drenckhan05}
W.~Drenckhan, S.~Cox, G.~Delaney, H.~Holste, D.~Weaire, and N.~Kern,
\newblock Rheology of ordered foams--on the way to Discrete Microfluidics,
\newblock Colloids Surf. A {\bf 263}(1-3), 52 -- 64 (2005).

\bibitem{guillen12}
V.~R. Guillen, M.~I. Romero, M.~da~Silveira~Carvalho, and V.~Alvarado,
\newblock Capillary-driven mobility control in macro emulsion flow in porous
  media,
\newblock Int. J. Multiphas. Flow {\bf 43}, 62--65 (2012).

\bibitem{raven06b}
J.-P. Raven and P.~Marmottant,
\newblock Periodic microfluidic bubbling oscillator : insight into the
  stability of two-phase microflows,
\newblock Phys. Rev. Lett. {\bf 97}, 154501 (2006).

\bibitem{green09}
T.~Green, P.~Grassia, L.~Lue, and B.~Embley,
\newblock Viscous froth model for a bubble staircase structure under rapid
  applied shear: An analysis of fast flowing foam,
\newblock Colloids Surf. A {\bf 348}, 49 -- 58 (2009).

\bibitem{brakke92}
K.~Brakke,
\newblock The Surface Evolver.,
\newblock Exp. Math. {\bf 1}, 141--165 (1992).

\bibitem{bruus}
H.~Bruus,
\newblock {\em Theoretical Microfluidics},
\newblock Oxford University Press, Oxford, 2008.

\bibitem{dollet10}
B.~Dollet and I.~Cantat,
\newblock Deformation of soap films pushed through tubes at high velocity,
\newblock J. Fluid Mech. {\bf 652}, 529--539 (2010).

\end{thebibliography}

\end{document}